\documentclass[a4paper]{article}\usepackage[]{graphicx}\usepackage[]{color}
\makeatletter
\def\maxwidth{ %
  \ifdim\Gin@nat@width>\linewidth
    \linewidth
  \else
    \Gin@nat@width
  \fi
}
\makeatother

\definecolor{fgcolor}{rgb}{0.345, 0.345, 0.345}

\usepackage{framed}
\makeatletter
 {\par\unskip\endMakeFramed%
 \at@end@of@kframe}
\makeatother

\definecolor{shadecolor}{rgb}{.97, .97, .97}
\definecolor{messagecolor}{rgb}{0, 0, 0}
\definecolor{warningcolor}{rgb}{1, 0, 1}
\definecolor{errorcolor}{rgb}{1, 0, 0}
\newenvironment{knitrout}{}{} 

\usepackage{alltt}
\usepackage{authblk} 
\usepackage{amsmath} 
\usepackage[comma]{natbib} 
\usepackage{enumitem} 
\usepackage{hyperref} 
\usepackage{url} 
\usepackage{relsize} 
\usepackage{comment} 
\usepackage{tabularx}
\usepackage[final]{changes} 
\usepackage[noframe]{showframe} 
\providecommand{\keywords}[1]{\textbf{\textit{Keywords:~}} #1}
\renewcommand{\a}{\alpha}
\renewcommand{\o}[1][]{\boldsymbol{o}_{#1}} 
\renewcommand{\d}[1][]{\boldsymbol{d}_{#1}}
\newcommand{\A}[1][]{\mathbf{A}_{#1}}
\newcommand{\B}[1][]{\mathbf{B}_{#1}}
\newcommand{\C}[1][]{\mathbf{C}_{#1}}
\newcommand{\D}[1][]{\mathbf{D}_{#1}}
\newcommand{\Y}[1][]{\mathbf{Y}_{#1}}
\newcommand{\Z}[1][]{\mathbf{Z}_{#1}}
\newcommand{\dstar}[1][]{\boldsymbol{d}_{#1}^{*}}
\newcommand{\Dstar}[1][]{\mathbf{D}_{#1}^{*}}
\newcommand{\Ystar}[1][]{\mathbf{Y}_{#1}^{*}}
\newcommand{\Zstar}[1][]{\mathbf{Z}_{#1}^{*}}
\newcommand{\Prod}[2][]{\mathlarger\prod_{#2}^{#1}} 
\newcommand{\one}[1]{\boldsymbol{1}\left\{#1\right\}}
\IfFileExists{upquote.sty}{\usepackage{upquote}}{}
\begin{document}

\title{
  \textbf{A Social Network Analysis of Articles on Social Network Analysis}
}
\author[1,2]{Clement Lee}
\author[1]{Darren J Wilkinson}
\affil[1]{School of Mathematics, Statistics and Physics, Newcastle University, UK}
\affil[2]{Open Lab, School of Computing, Newcastle University, UK}
\maketitle

\begin{abstract}
  A collection of articles on the statistical modelling and inference of social networks is analysed in a network fashion. The references of these articles are used to construct a citation network data set, which is almost a directed acyclic graph because only existing articles can be cited. A mixed membership stochastic block model is then applied to this data set to soft cluster the articles. The results obtained from a Gibbs sampler give us insights into the influence and the categorisation of these articles.
\end{abstract}
\\

\noindent
\textbf{\textit{Correspondence:~}}clement.lee@newcastle.ac.uk\\
Open Lab, Level 1, Urban Sciences Building, Newcastle University, Newcastle upon Tyne, NE4 5TG, United Kingdom

\clearpage
\section{Introduction} \label{sect.intro}
Social network analysis can be applied to a wide range of topics and data. One of its applications is in bibliometrics, where networks concerning academic publications are being constructed and then analysed in a quantitative way. While different networks could be constructed from the same data, it is networks with authors as the nodes that received more attention. For example, \cite{newman01a} and \cite{newman04a} analysed scientific papers from the perspective of (co-)authors, with a focus on comparing the coauthorship networks of the subjects studied. \cite{newman01c} suggested that such coauthorship networks form ``small worlds'', which is one common phenomenon in various kind of networks \citep{ws98}. \cite{jj16} constructed the coauthorship network of statisticians, by looking into all research articles in four top statistics journals over 10 years, and constructed the citation network based on the same data set.

While citation networks might have received less attention, there have been influential and interesting analyses on them. For example, \cite{price76} investigated the (in-)degree of articles in citation networks, and proposed the idea of ``cumulative advantage process'' where success breeds success, which later became the preferential attachment model by the highly cited \cite{ba99a}. \cite{vcf16} investigated the citation exchange between statistics \textit{journals}, with a focus on aiding the comparisons of journal rankings.

One issue regarding the application of social network analysis to citation networks is that they are not true social networks, as the authors of the citing and the cited articles do not necessarily know each other \citep{newman01a}. While this issue makes a model that concerns the generating mechanism of a network less appropriate for such data, a model that concerns summarising or clustering a network may still uncover interesting structures or characteristics of a citation network. This will be the focus of this article.

Clustering articles in the literature via a modelling approach is useful as the results can be compared with the prior insights into the literature exhibited in these articles, in particular the review articles. One potential issue of treating all citations in an article equally in a quantitative analysis is that the importance of different citations to the article, which there is no automatic way of differentiating, is not accounted for. However, we argue that the citations as a whole should give some sense of how the authors position their article, as well as what works in the literature are deemed relevant to them.

Regarding the statistical modelling and inference of social networks, there are (at least) three main groups of models, namely the generative models including the small-world model \citep{ws98} and the preferential attachment model \citep{ba99a}, the exponential random graph models (ERGMs), and the latent models including the latent space model \citep{hrh02} and the stochastic block models \citep{hll83}. While comprehensive reviews for the majority of these three groups can be found in \cite{hks12}, \cite{ab02a}, and \cite{mr14}, respectively, they will also be briefly reviewed in Section \ref{sect.review}. The main reason is that all the articles cited \textit{in this article} will form the node list of the data, to which our proposed model is applied. Subsequently, the references of these articles become the edge list, and the clustering analysis presented in this article can be viewed as a literature review, not in the traditional sense, but in a network fashion faithful to the articles being reviewed.

Using the articles cited (and their references) as the data comes with two common criticisms of non-random sampling: selection bias and confirmation bias. Regarding the selective nature of the articles cited, we argue that we want to keep the articles analysed limited and relevant so that insights can be drawn on them individually in the application. This is in contrast to the more systematic approach of collecting the data by, for example, \cite{vcf16} and \cite{jj16}. Regarding that the results are confirming or perhaps influencing the categorisation in Section \ref{sect.review}, we argue that we have attempted to include all of the most relevant articles across the domain of interest, and since our focus is on the modelling approach of clustering the articles, the results shed new light on how they can be categorised in a less subjective way. Furthermore, the proposed mixed membership stochastic block model (MMSBM) provides a soft clustering of the articles, thus allowing us to realise how articles are positioned between topics, and (re-)discover interdisciplinary works.

The data set presented is very transparent as one can look up all the references in the bibliography and recompile the same data set. However, data cleaning has already been done to ensure its quality, especially in terms of updating obsolete references to ``re-link'' the supposedly linked articles. Furthermore, the data set can always be augmented or modified by the reader before (re-)applying the proposed model or any other relevant social network models. Ultimately, this article and the data set can serve as a starting point for anyone who is entering the field and would like to know how different articles they come across position in the literature.

Comparing to the original MMSBM \citep{abfx08}, the proposed model halves the number of latent variables, potentially reducing the statistical uncertainty, and simultaneously enables the \textit{topological} order of the articles to be inferred, which comes in the model as a parameter. Finally, it can be applied not only to a citation network, but any network that is a directed acyclic graph (DAG), such as software dependencies.

The rest of this article is as follows: Section \ref{sect.review} briefly reviews selected articles on social network analysis. These articles will form the node list of the data, on which an exploratory analysis is carried out in Section \ref{sect.eda} after data cleaning. The proposed model is introduced in Section \ref{sect.model}, with its likelihood derived and inference algorithm outlined in Section \ref{sect.lik_inf}. The application to the citation data is presented in Section \ref{sect.app}. Section \ref{sect.discuss} concludes the article.

\section{Literature review and data} \label{sect.review}

In this section we give a brief review of the selected relevant articles, with a focus on, but not limited to, the statistical modelling and inference of social networks. All of the articles cited in Section~\ref{sect.intro} are included for the sake of completeness. This review is by no means a replacement of any of the reviews mentioned in Section~\ref{sect.review_review}, but a presentation of part of our data, namely the nodes with non-zero out-degrees in the resulting citation network. Furthermore, the categorisation of the articles here can be seen as a hard clustering of the data, done manually using our prior knowledge of the literature.

\subsection{Generative models}
The first main group of articles concerns models which can generate a network via a few simple rules. These models are called the ``pseudo-dynamic'' models by \cite{gzfa10}. The first of this kind is the Erd\"{o}s-R\'{e}nyi model \citep{er59,er60}, also known as the Bernoulli random graph model, in which any pair of nodes (a dyad) has the same probability of having an edge, and is independent of all other dyads. \cite{milgram67} proposed and \cite{ws98} popularised the small-world model, in which some edges of a Bernoulli random graph are ``rewired'' to achieve the small-world phenomenon, which was also analysed by, for example, \cite{wc03} and \cite{kleinberg08}, and observed in scientific coauthorship networks \citep{newman01c}.

Another commonly observed phenomenon in realistic networks is that they are scale-free, meaning that the degree distribution approximately follows a power law. \cite{price76} proposed a mechanism, which was called cumulative advantage and later coined preferential attachment \citep{ba99a}, to generate such a network. The preferential attachment model was extended or further studied by \cite{ajb00}, \cite{krr01}, \cite{vpv02a}, \cite{wc03}, \cite{rdp04}, \cite{hthl07}, and \cite{varga15}, while the more general scale-free phenomenon was studied by \cite{ajb99}, \cite{fff99}, \cite{newman01a}, and \cite{swm05}. \cite{njaj05} proposed a model which generates a network with a group structure.

It is also possible to generate a network given the degree sequence, such as the configuration model \citep{xl07}. Related is the study by \cite{mr95} on the properties of random graphs with a given degree sequence. Other miscellaneous models of generating networks include \cite{krrstu00b} and \cite{kl11a}. Finally, \cite{nsw01} studied and compared the clustering properties of the networks generated by various models aforementioned.




\subsection{Exponential random graph models}
Another prominent group of models are the exponential random graph models (ERGMs). Forerunners include \cite{hl81}, \cite{fw81a}, \cite{fs86} and \cite{strauss86}, and further developments include the trilogy of \cite{wp96}, \cite{pw99} and \cite{rpw99}, and others such as \cite{vsz04}, \cite{hh06} and \cite{fh13}. In its most basic form, the ERGMs on topological space is analogous to the exponential family distributions on Euclidean space. Essentially, the log-likelihood is, up to a constant, a linear combination of various summary statistics of the given graph. A summary of the formulation and application of ERGMs for social networks is provided by \cite{rpkl07}.

One major issue with the ERGMs is that the exact likelihood is difficult to compute \citep{hh06}, thus impeding inference. Consequently, methods based on pseudo-likelihood, constrained likelihood or conditional likelihood have been proposed, such as \cite{si90}, \cite{gt92}, and \cite{sd17}. Markov chain Monte Carlo (MCMC) is used by, for example, \cite{snijders02} and \cite{hh06}, with further advances by \cite{cf11} and \cite{cf13} based on the exchange algorithm by \cite{mgm06}. Extensions of the MCMC approach have been made by \cite{krp10} and \cite{krwp13} to allow for missing data, as well as by \cite{tfck16} and \cite{sk16} to incorporate random effects and multiple levels, respectively. \cite{bfm17} proposed to combine the best of both worlds in a Bayesian framework by exploiting the computational efficiency of the pseudo-likelihood approach while correcting for the resulting MCMC algorithm.

While the inferential algorithms in the aforementioned articles have circumvented the computational difficulty, they are not able to eradicate another major issue, which is model degeneracy and poor fit \citep{handcock03a}. This means that the probability mass concentrates on only a few graph configurations, usually including the empty graph (no nodes connected) and/or the complete graph (any pair of nodes connected). Simulation or sampling using, for example, the maximum likelihood estimates then generates unrealistic graphs, thus undermining the usefulness of the model. \cite{hgh08} provided a systematic examination of the goodness-of-fit and degeneracy of then existing models.

A major factor of the usefulness of an ERGM is the specification of the summary statistics or configurations of the graph. Therefore, efforts have been made on finding or proposing new specifications to improve the model fit, such as \cite{sprh06}, which are in turn reviewed by \cite{rswhp07}. \cite{sh15} and \cite{nmd17} proposed new specifications to address local dependence and homophily, respectively, which are features exhibited by real networks, while \cite{tk17} replaced the linear statistics by smooth functional components.

While the majority of articles on ERGMs focus on issues with modelling, specifications and inference, there have also been computational and theoretical developments. For example, \cite{hhbgm08b} developed the \texttt{R} package \texttt{ergm} for fitting and simulating from ERGMs. Related is the package \texttt{statnet} \citep{hhbgm08a} for social network analysis in general. \cite{rfz09}, \cite{cd13} and \cite{sr13} examined maximum likelihood estimation for ERGMs from a theoretical perspective.

\subsection{Community detection and latent models}
Clustering of relational data is also called community detection. The community or group membership of a node in the network is not defined by the attributes of or the content associated the node itself, but by how it is connected to other nodes. Nodes are closely connected within each group while connections between different groups are much weaker. Forerunners in clustering include \cite{lw71}, \cite{bba75} and \cite{wbb76}.

\cite{bba75} is one example of the algorithmic approach of community detection. Other community detection algorithms are usually based on the optimisation of certain summary measures, such as centrality \citep{gn02} and modularity \citep{newman06b,bc09a}. A comprehensive review of community detection algorithms can be found in \cite{fortunato10}.

The modelling approach of \cite{wbb76} led to the family of stochastic block models (SBMs). While other models take networks as ``given'', a SBM assumes for each node a latent variable which represents the group membership, and the edge inclusion probability of any two nodes depend solely on their respective memberships. Fitting the SBM essentially uncovers the latent structure of the network. Fundamental results regarding modelling and inference were established by \cite{hll83}, \cite{sn97} and \cite{ns01}, while extensions have been made to multilevel data \citep{fmw85,sstm16,bdlb17}, directed graphs \citep{ww87}, valued graphs \citep{mrv10}, networks with textual edges \citep{blz16}, incorporating temporal dynamics \citep{len17,mm17}, to name a few.

Apart from the extensions above, there have been other developments on the application and properties of the SBMs. For example, numerous attempts have been made on solving the issue with the number of groups, which is prespecified in the original SBM. There include, for example, \cite{mmfh13}, \cite{cl15}, \cite{yan16}, \cite{hkk16}, and \cite{peixoto18a}. Another development concerns the relationship of the SBMs with other methods and models. \cite{newman16} and \cite{vv18} established connections between modularity optimisation and inference for SBMs, using frequentist and Bayesian approaches, respectively. Finally, although SBMs are prevalent in the literature, there are also alternatives proposed. For example, \cite{dpr08} and \cite{vhs13} considered mixture models for model-based clustering of large networks.

Apart from the prespecified number of groups, another issue with the original SBM is the hard clustering, that is, each node can belong to one latent group only. \cite{abfx08} combined the SBM with latent Dirichlet allocation (LDA) \citep{bnj03} to form the mixed membership stochastic block model (MMSBM). Since then there have been extensions based on the MMSBM, such as \cite{gmgfb12}, \cite{stj14}, \cite{fxc16} and \cite{law16}. The MMSBM will be central to our modelling approach in Section \ref{sect.model}.

Related to SBMs are latent space models, proposed by \cite{hrh02}, and applied by \cite{hrt07}. While SBMs uncover the latent structure of the nodes in the topological space, the latent space models assume a latent projection of the nodes in a $d$-dimensional Euclidean space, and that the observed network arises from how they are positioned in this latent space. \cite{abfx08} noted the connection between the latent space model and the MMSBM. Other studies on latent social structure which do not rely on the model by \cite{hrh02} include \cite{sm05}, \cite{bpdfd08}, \cite{tl09}, \cite{ml10} and \cite{dd16}.

\subsection{Reviews and themes across groups} \label{sect.review_review}
Various reviews in the literature on different aspects of social network analysis came with different foci on the three main group of articles outlined above. Quite a few of them focus on one main group, such as \cite{ab02a} and \cite{newman03a} for mostly generative models, and \cite{mr14} for the latent models and SBMs. In these three reviews there have been no or very little mention of ERGMs, which are however reviewed extensively by \cite{hks12}, with a focus on computational and inferential aspects. Regarding reviewing mostly two of the three main groups of articles, \cite{fortunato10} excluded ERGMs while \cite{snijders11} and \cite{fienberg12} excluded the generative models. Reviews which span over all three groups include \cite{gzfa10}, \cite{snijders11}, \cite{jackson11}, \cite{oo14} and \cite{channarond15}, the last of which also cited fellow reviews including \cite{fortunato10}, \cite{gzfa10}, \cite{snijders11} and \cite{mr14}. Finally, the review by \cite{ke05} does not largely fall into any of the three groups, as its focus is on networks and epidemics. It is included here (and therefore in our data) because of its relevance, as well as its role as a starting point into the literature on networks and epidemics outside the scope of our analysis.

We single out the reviews because it is usually difficult to put one in a single group due to their very nature. This is exhibited by their different positioning aforementioned. It would be useful to quantify the proportion of a review in each of the main groups, which is exactly what a mixed-membership model does. Therefore, in our application in Section~\ref{sect.app}, we will visualise their positioning among the three main groups. These results will be useful signposts for anybody wanting to study the literature but not knowing where to start.


\subsubsection{Inference}
There are at least three aspects of statistical inference for network models, which are shared by articles across the three main groups and worth mentioning here. The first is inferring or predicting the unobserved edges in a network given the observed ones, under names such as link prediction, missing data, or network completion. They include the previously mentioned \cite{krp10} for ERGMs, \cite{kl11b}, and \cite{zwlz17}. The second aspect is the goodness-of-fit and comparison of networks models, examined by the likes of \cite{bks06}, \cite{hgh08}, and \cite{fmbfj14}. Finally, inference on network statistics are considered by \cite{fmbfj14} and \cite{gd17}.

\subsubsection{Network dynamics}
The topic of (temporal) dynamics of networks is not the main focus of the articles reviewed here, as we must limit our scope. However, it is natural that it is touched upon by some of them, which are therefore worth mentioning. \cite{pss15}, \cite{dd16} and \cite{len17} incorporated temporal dynamics, in preferential attachment networks, latent space models and SBMs, respectively. \cite{bhkl06} considered growth of communities in social networks, while \cite{thompson13} applied network dynamics in spatial networks. For a brief review on network dynamics, see, for example, \cite{snijders01}.

\subsection{Miscellaneous}
Several influential articles are worthy of inclusion but cannot be placed in a single group, precisely because they are highly cited by the references in this article and in general. They include \cite{bp98}, who introduced the PageRank centrality, \cite{dlr77}, who proposed the Expectation-Maximisation algorithm, and \cite{besag74,besag75}, which are seminal articles on the (spatial) modelling of lattice data.

Several other articles cannot be comfortably put into one of the three main groups above because of their uniqueness. \cite{pa93} presented the Medici family network, which a classic data set used in social network analysis \citep{pw99,rpkl07,cf11}. \cite{xn11} provided an asymptotic analysis of maximum likelihood estimation in the framework of Markov networks for relational learning. \cite{bbrsf16} surveyed algorithms for reordering adjacency matrices of networks. \cite{jj16} and \cite{vcf16} are case studies on the citation network of statisticians and statistics journals, respectively. \cite{eg18} established a connection between the ERGMs and the SBMs. Nevertheless, they are all included in our data as they have cited and/or been cited by others in the main groups.

The Proceedings of the National Academy of Sciences published a special issue on ``Mapping Knowledge Domains'' \citep{sb04}, in which the articles published in the same journal were analysed from different perspectives. Among all the articles in this special issue, \cite{bmg04}, \cite{menczer04}, \cite{newman04a}, \cite{gs04b} and \cite{efl04} are more relevant to social network analysis than the rest, and therefore included in our data.


As the references included so far are journal and conference articles, it is notable that books (and theses) are excluded. This is due to the difficulty of compiling the volume of references in the books, many of which may not be relevant to social network analysis. Note that if a book has ever been cited by an article in our data, it will still become part of the data, as a receiving node in the edge list. It will however be excluded from our application due to the information asymmetry, as we do not know which articles are cited by this book.

The difficulty of categorising the above miscellaneous articles echoes that for the reviews in Section~\ref{sect.review_review}, and exemplifies the issue with hard clustering. Therefore we argue for the soft clustering approach of the MMSBM \citep{abfx08} in our model and application.



\section{Data cleaning and exploratory analysis} \label{sect.eda}

As previously mentioned, the references in this article, which are called citing articles hereafter, are used to construct our data set. Using all of their references initially, which are called cited articles hereafter, the edge list of the graph, whose nodes are the union of the citing and cited articles, can be compiled. In total there are 7,010 edges and 4,026 nodes, including 135 citing articles. While the general rule applies that if article A cites article B, article B will not cite article A, there are few exceptions due to time proximity between the articles in question, or the overlap in one or more of the authors of the two articles, or the nature that we update the attributes of the articles to their published version in the data. For each such pair of articles, the directed edge from the earlier article to the latter article, in terms of publication year (and issue/month), is removed from the data. In cases where there is a tie or no apparent ways of differentiating them chronologically, either of the two directed edges is removed randomly with equal probability. These removed directed edges are listed in Table \ref{table.antiedges}. While \cite{sprh06} actually cited an earlier version of the \verb'R' package \verb'statnet', we take the discretion of treating this as an edge from \cite{sprh06} to \cite{hhbgm08a}, the article which introduced the same package in the Journal of Statistical Software. 

\begin{table}[!htbp]
  \centering
  \begin{tabularx}{\textwidth}{|X|X|}
    \hline
    From & To \\
    \hline
    \cite{bba75} & \cite{wbb76} \\
    \cite{er59} & \cite{er60} \\
    \cite{fmw85} & \cite{fs86} \\
    \cite{fs86} & \cite{strauss86} \\
    \cite{fw81a} & \cite{hl81} \\
    \cite{rfz09} & \cite{gzfa10} \\
    \cite{handcock03a} & \cite{sprh06} \\
    \cite{hh06} & \cite{sprh06} \\
    \cite{hhbgm08a} & \cite{hhbgm08b} \\
    \cite{wp96} & \cite{pw99} \\
    \cite{sprh06} & \cite{hhbgm08a} \\
    \cite{menczer04} & \cite{bmg04} \\
    \cite{hh06} & \citet{hgh08} \\
    \cite{newman01c} & \cite{nsw01} \\
    \cite{newman01a} & \cite{nsw01} \\
    \hline
  \end{tabularx}
  \caption{Directed edges that cause circular citations and thus are removed.}
  \label{table.antiedges}
\end{table}

The resulting graph is very sparse with a density $6995/(4026\times 4025/2.0)=0.086$\%. The division by 2 is due to the assumption that we are working with a directed acyclic graph (DAG), which is true as there are no longer loops found in the resulting graph, after removing the edges that cause circular citations. Such a low density is partly due to the incompleteness of information, as we do not know if the cited articles here are in turn referencing the citing articles. If we consider the sub-graph with only the citing articles as nodes, the density becomes $1118/(135\times 134/2.0)=12.36$\%. Due to a higher density and the completeness of information of the edges, all analyses performed hereafter are based on this sub-graph, unless otherwise stated. It will be simply called the graph of citing articles, of which the network plot is shown in Figure \ref{fig.plot_network_subgraph}. 

\begin{knitrout}
\definecolor{shadecolor}{rgb}{0.969, 0.969, 0.969}\color{fgcolor}\begin{figure}[htbp!]

{\centering \includegraphics[width=0.75\linewidth]{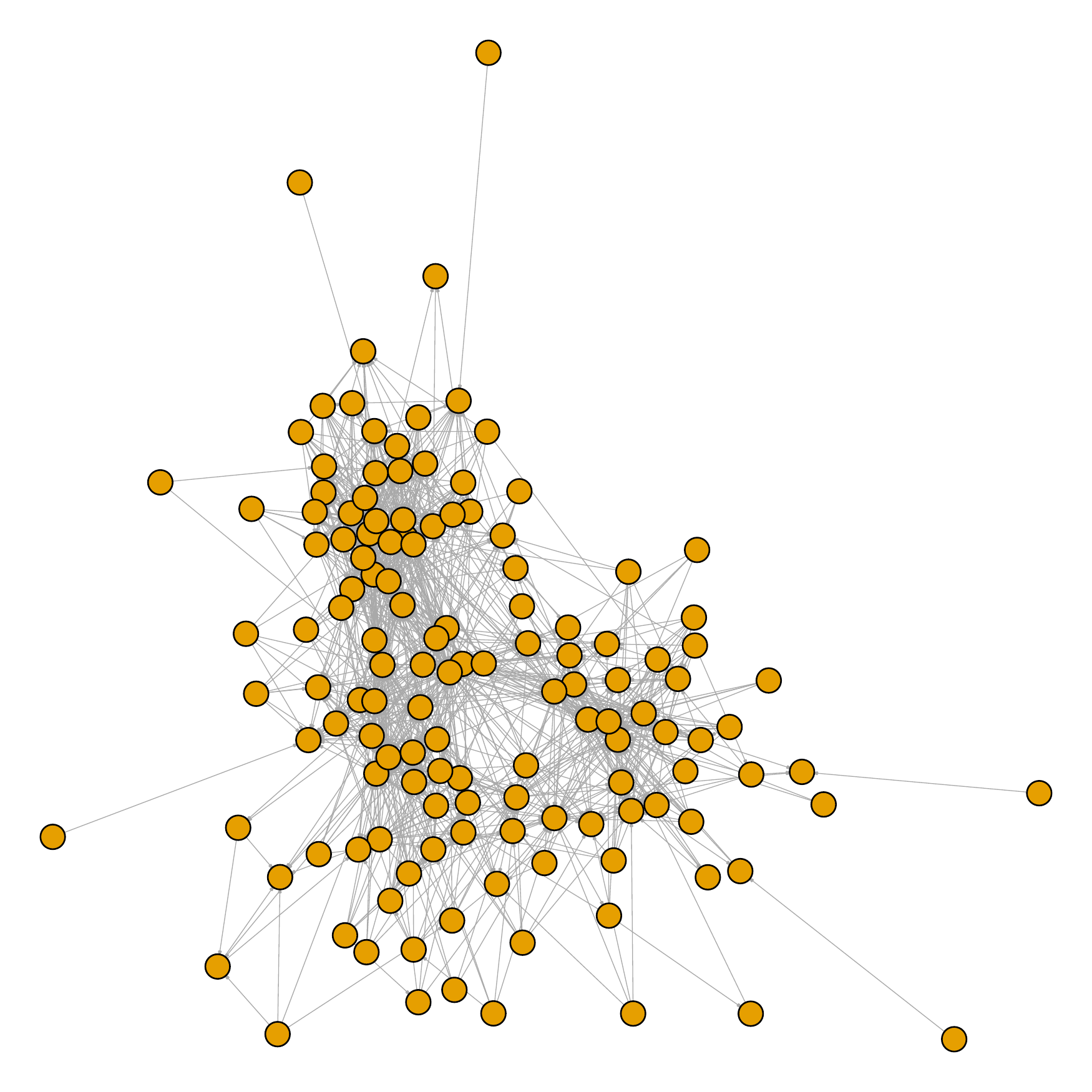} 

}

\caption[Network plot of graph of citing articles]{Network plot of graph of citing articles.}\label{fig.plot_network_subgraph}
\end{figure}

\end{knitrout}


Before clustering the articles via a modelling approach, we first use the spinglass and the walktrap algorithms to perform community detection. The adjacency matrix of the graph of citing articles, sorted according to the results of each of the two algorithms, is plotted in Figure \ref{fig.plot_adjacency_subgraph}. It is quite visible that either algorithm detects three main groups. If the two sets of results are matched manually, we can see a large intersection for each pair of groups, as shown in Figure \ref{fig.plot_venn_subgraph}. What is even more encouraging is that each intersection also corresponds roughly to a main group of articles reviewed in Section \ref{sect.review}.

\begin{knitrout}
\definecolor{shadecolor}{rgb}{0.969, 0.969, 0.969}\color{fgcolor}\begin{figure}[htbp!]

{\centering \includegraphics[width=0.3\linewidth]{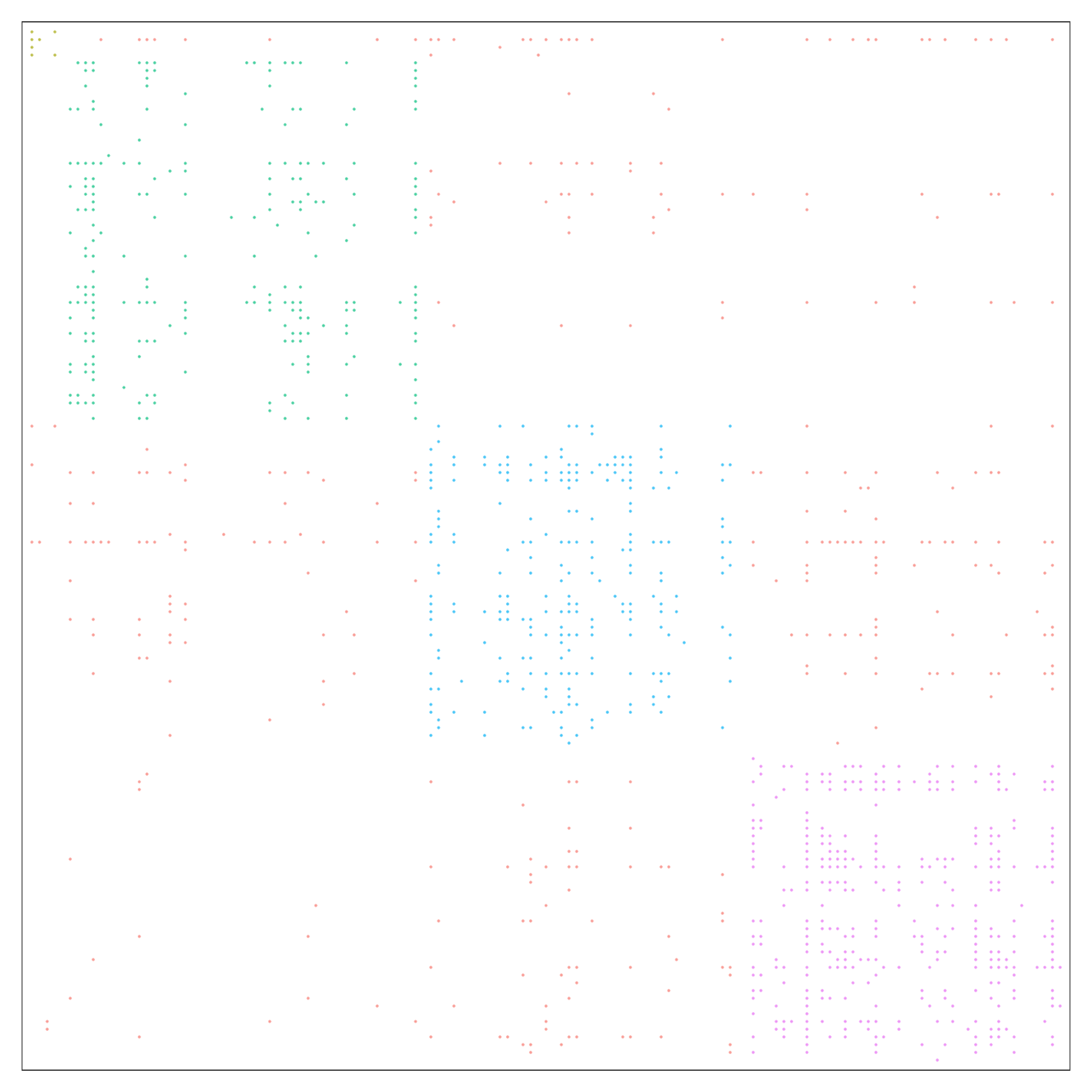} 
\includegraphics[width=0.3\linewidth]{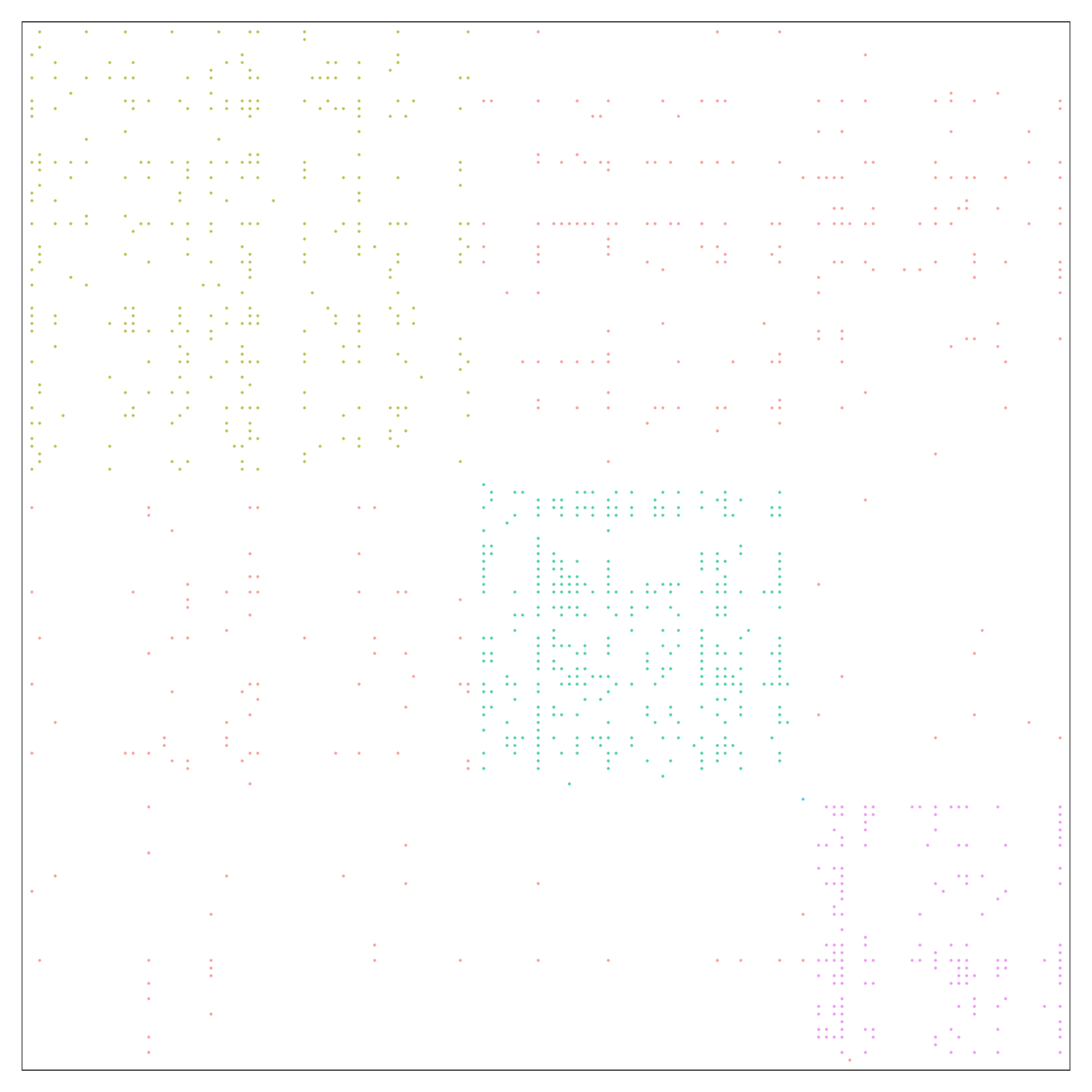} 
\includegraphics[width=0.3\linewidth]{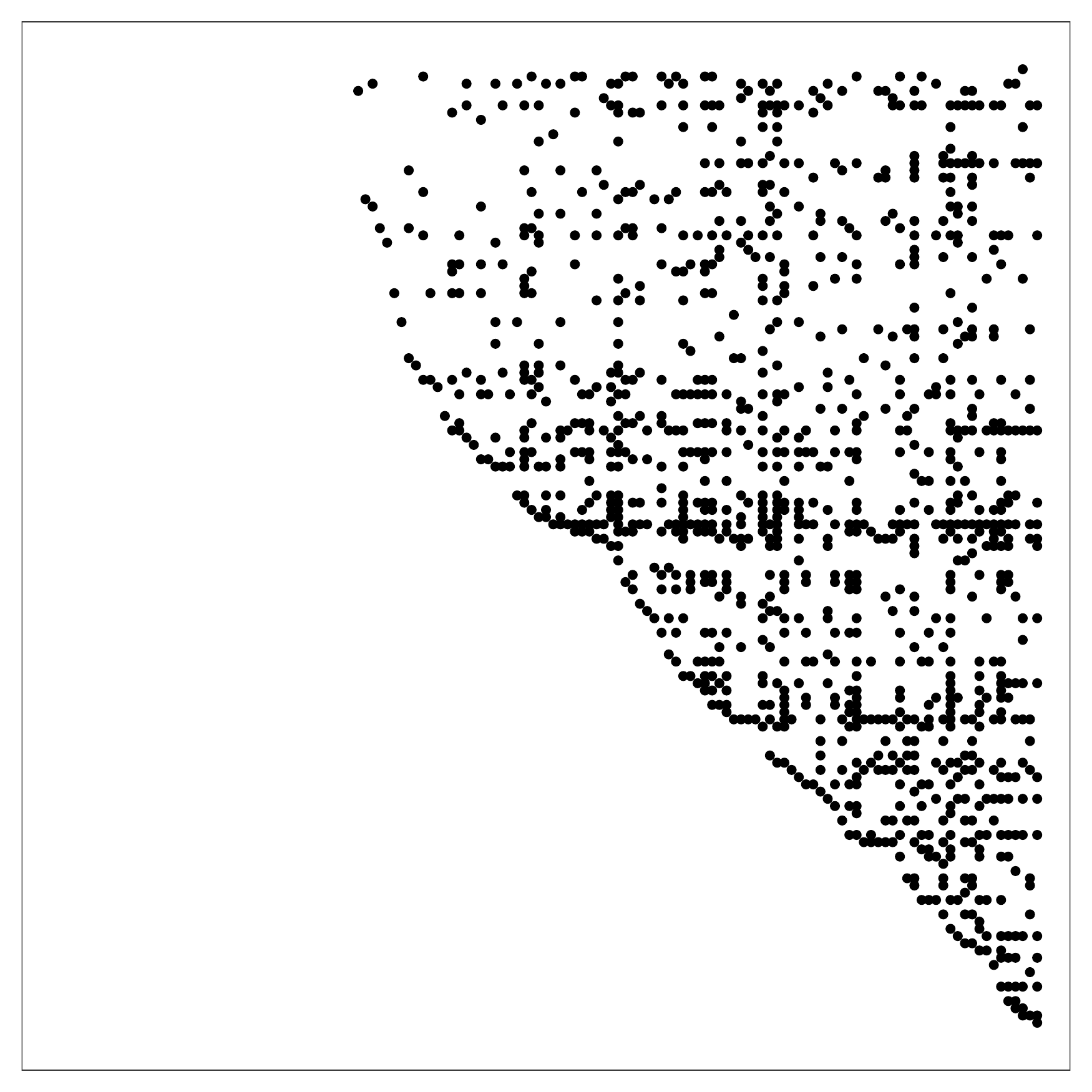} 

}

\caption[Adjacency matrix of the graph of citing articles, sorted according to community detection by the spinglass algorithm (left) and the walktrap algorithm (middle), and according to one topological order (right)]{Adjacency matrix of the graph of citing articles, sorted according to community detection by the spinglass algorithm (left) and the walktrap algorithm (middle), and according to one topological order (right).}\label{fig.plot_adjacency_subgraph}
\end{figure}

\end{knitrout}

\begin{knitrout}
\definecolor{shadecolor}{rgb}{0.969, 0.969, 0.969}\color{fgcolor}\begin{figure}[htbp!]

{\centering \includegraphics[width=0.35\linewidth]{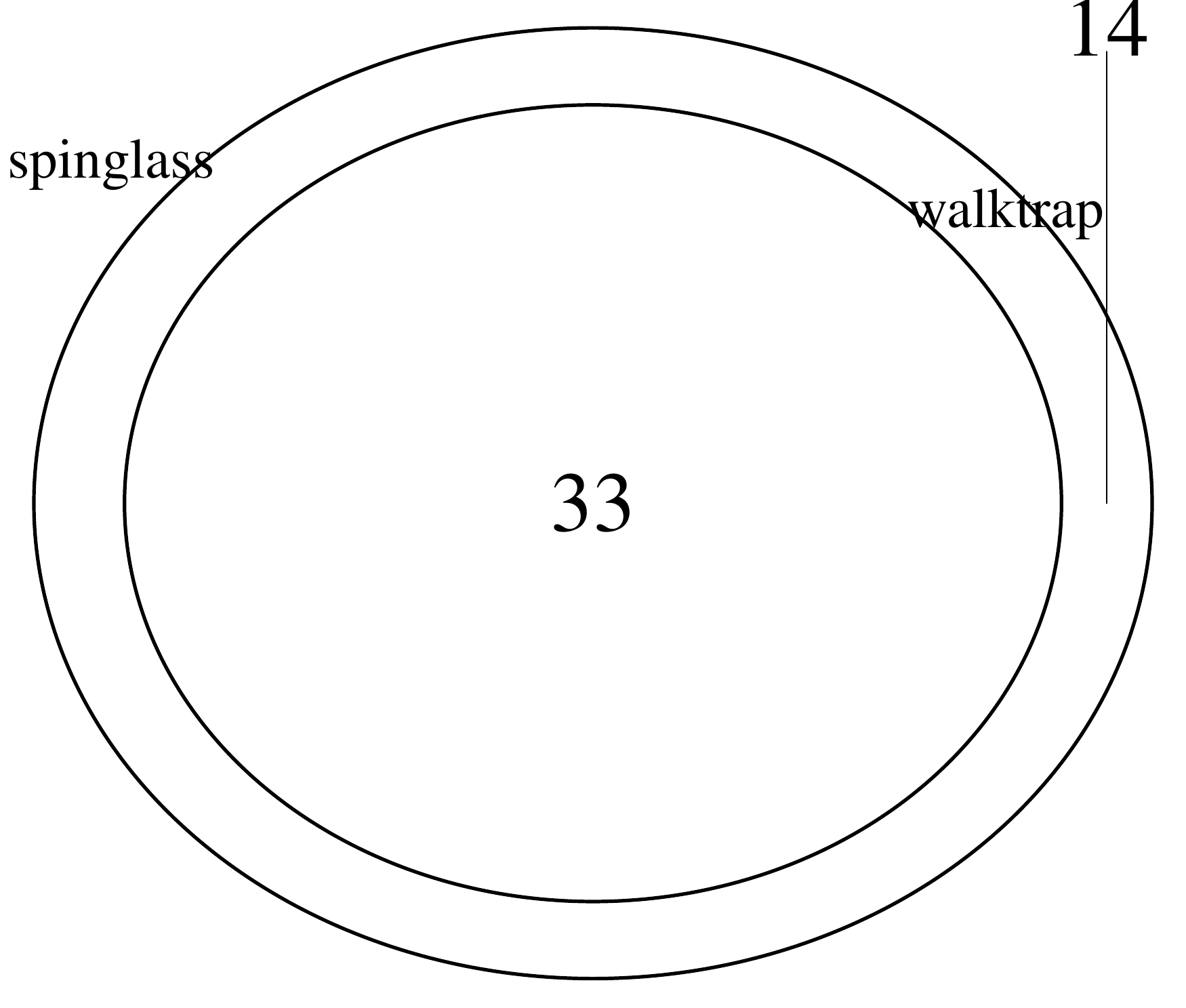} 
\includegraphics[width=0.35\linewidth]{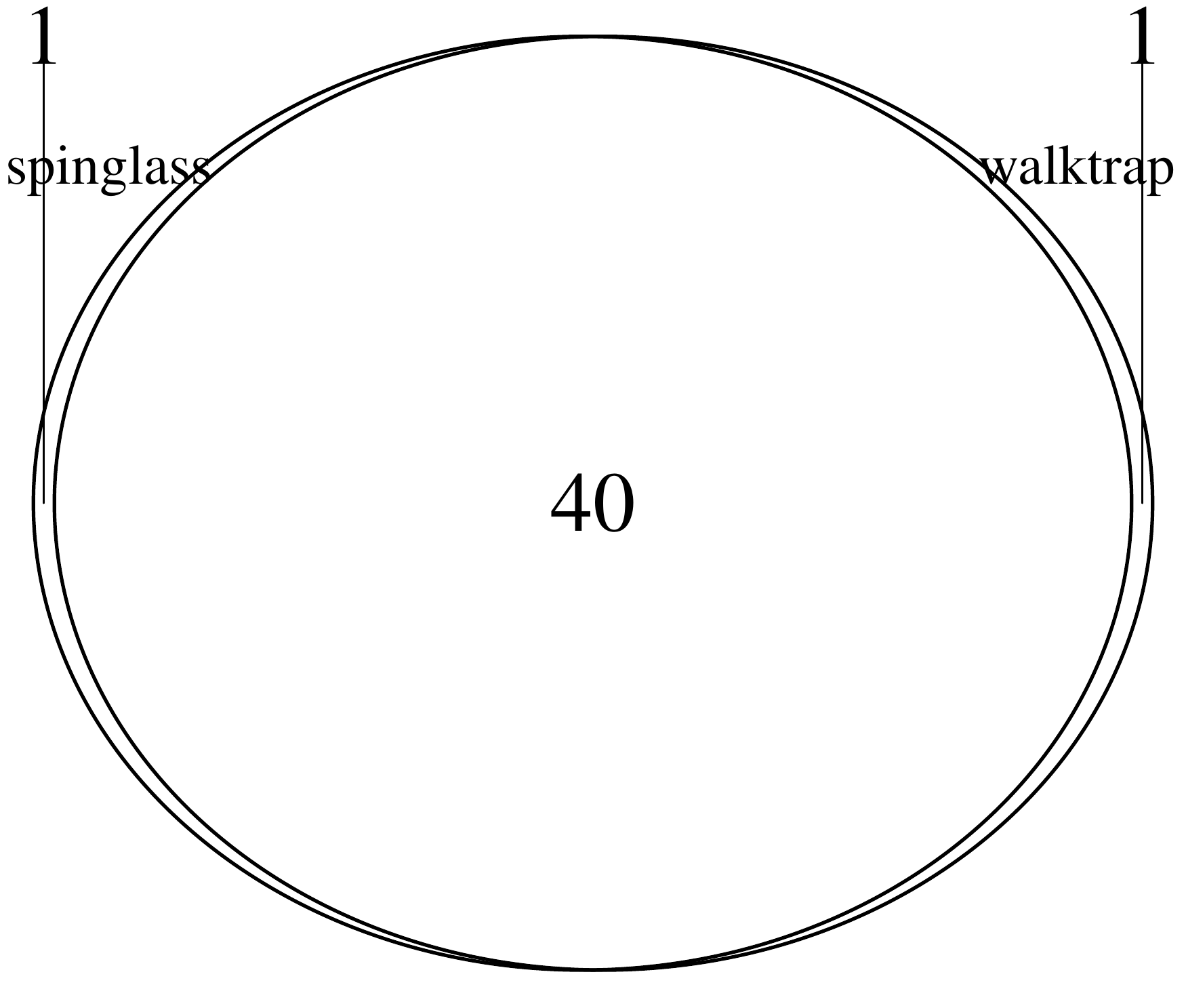} 
\includegraphics[width=0.35\linewidth]{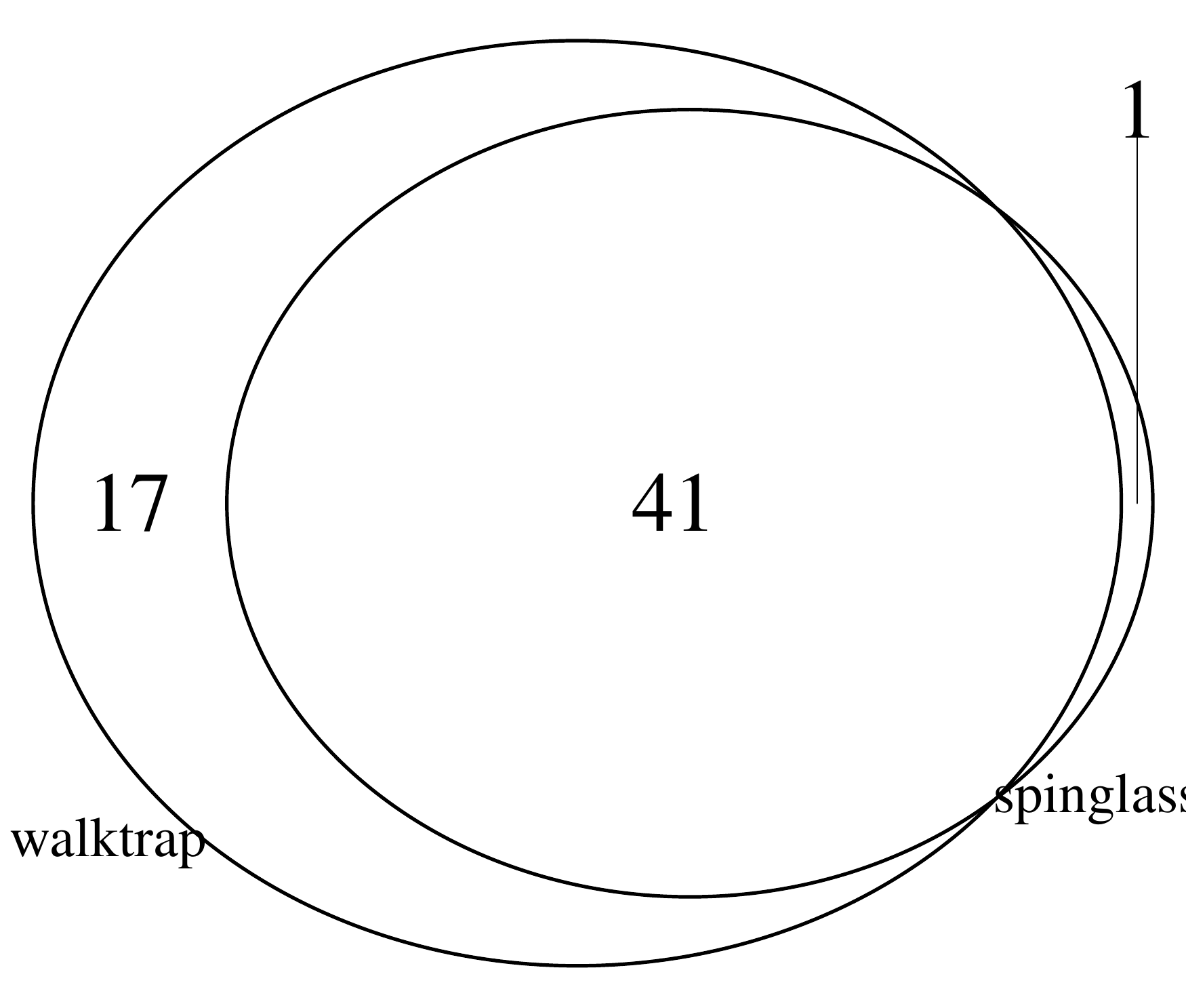} 

}

\caption[Venn diagram of the generative (top), ERGM (middle) and latent (bottom) groups detected by the spinglass and walktrap algorithms]{Venn diagram of the generative (top), ERGM (middle) and latent (bottom) groups detected by the spinglass and walktrap algorithms.}\label{fig.plot_venn_subgraph}
\end{figure}

\end{knitrout}

One final exploratory analysis concerns the topological order of the articles, as the graph is now a DAG. If article A is before article B topologically, it means that article B will not be citing article A. Therefore, the more recent articles are more likely to be at the front of the toplogical order. The adjacency matrix, sorted according to one topological order, is plotted on the right of Figure \ref{fig.plot_adjacency_subgraph}. However, such order is not unique and is not given in the data. Therefore it will be treated as a parameter in our model, and its posterior distribution will be compared with the chronological order of the articles.

\section{Model} \label{sect.model}
In this section we present a mixed membership stochastic block model (MMSBM) of $K$ latent groups, for a DAG, denoted by $\mathcal{G}=\left(\mathcal{N},\mathcal{E}\right)$, where $\mathcal{N}$ is the node set of size $n$, and $\mathcal{E}$ is the edge list of size $m$. The corresponding $n\times n$ adjacency matrix is denoted by $\Y$, which means $\Y[rs]=1$ when there is a directed edge from node $r$ to $s$, $\Y[rs]=0$ otherwise, where $\mathbf{M}_{rs}$ represents the $(r,s)$\textit{-th} element of matrix $\mathbf{M}$. Self-loops are not allowed and $\Y[rr]~\left(r=1,2,\ldots,n\right)$ is set to 0. As $\mathcal{G}$ is a DAG, we introduce the topological order of the nodes, denoted by the $n$-vector $\o$, which is now a parameter in the model. Using $\o$, we define the reordered adjacency matrix $\Ystar:=\Y\left(\o\right)$ such that $\Ystar[pq]=\Y[{\o[p]\o[q]}]$ for $1\leq p,q\leq n$. The nature of $\o$ as the topological order ensures that $\Ystar$ is upper triangular. Similarly, we refer $\mathbf{M}^{*}:=\mathbf{M}\left(\o\right)$ to be the reordered matrix for any $n\times n$ matrix $\mathbf{M}$ hereafter. Essentially, $\Ystar$ is such that $\Ystar={\Ystar}^{+}$, where $\mathbf{M}^{+}$ is the upper triangular matrix of $\mathbf{M}$.

The essence of the MMSBM is that a node can belong to different groups when interacting with different nodes. Therefore we introduce the $n\times n$ matrix $\Z$, in which $\Z[rs]$ represents the group node $r$ is in \textit{when interacting with node $s$}. As self-interactions are not considered in the model, $\Z[rr]~(r=1,2,\ldots,n)$ is fixed and set to $-1$ for convenience. For $r\neq s$, $\Z[rs]$ is a latent variable with state space $\{1,2,\ldots,K\}$, assumed to be independent of $\Z[pq]$ for $(p,q)\neq(r,s)$, and has a multinomial distribution with probabilities $\d[r]$ \textit{apriori}, where $\d[r]$ is a (column) vector of length $K$. As a node has to be in one of the $K$ groups when interacting with another node, the membership probabilities in $\d[r]$ has to sum to 1, which means $\d[r]^{T}\boldsymbol{1}_{r}=1$, where $\mathbf{1}_{l}$ is an $l$-vector of $1$'s. We also define $\D:=\left(\d[1]~\d[2]~\cdots~\d[n]\right)^T$ as the $n\times K$ matrix of membership probabilities, so that $\D[ri]$ is the $i$\textit{-th} element of $\d[r]$ for $i=1,2,\ldots,K$. This means that $\d[r]^{T}\boldsymbol{1}_{r}=1$ is equivalent to $\sum_{i=1}^{K}\D[ri]=1$. Finally, we denote the reordered $\D$ according to $\o$ by $\Dstar$, such that $\Dstar[pi]=\D[{\o[p]i}]$ for $p=1,2,\ldots,n$ and $i=1,2,\ldots,K$. Essentially, the $\o[p]$\textit{-th} row of $\D$ is equivalent to the $p$\textit{-th} row of $\Dstar$, the transpose of which is denoted by $\dstar[p]$, for consistency with how $\d[r]$ is defined above. 

In order to describe the generation of the edges of $\mathcal{G}$ according to the group memberships of the pair of nodes concerned, a $K\times K$ matrix, denoted by $\C$, is introduced. For $1\leq i,j\leq K$, $\C[ij]\in[0,1]$ and represents the probability of occurrence of a directed edge from a node in group $i$ to a node in group $j$. Now, consider two nodes $r$ and $s$, and assume $r$ is ahead of $s$ topologically, without loss of generality. If we denote their positions in $\o$ by $p$ and $q$, respectively, such that $r=\o[p]$ and $s=\o[q]$, we have $p<q$. The latent variable $\Z[rs](=\Zstar[pq])$ is in the upper triangle of $\Zstar$, and represents the group membership of node $r$ in its interaction with node $s$. Similarly, $\Z[sr](=\Zstar[qp])$ is in the lower triangle of $\Zstar$, and represents the group membership of node $s$ in the same interaction concerned. On one hand, $\Y[sr]$ is equivalent to $\Ystar[qp]$, which is in the lower triangle of $\Ystar$ and is therefore 0. In the context of citation networks, an article at the back of the topological order cannot cite an article at the front of the topological order. On the other hand, $\Y[rs]$, or $\Ystar[pq]$ equivalently, follows the Bernoulli distribution with success probability $\C[{\Z[rs]\Z[sr]}]$. In the same context, this means whether article $r$ cites article $s$ depends on their specific group memberships when interacting \textit{and} the resultant group-to-group probability given by $\C$.

A few things need to be noted before deriving the likelihood in the next section. Firstly, the dyads in $\mathcal{G}$ are not independent marginally, but conditionally given the memberships of the nodes, or essentially $\Z$. Secondly, while the likelihood requires only half of the dyads, that is, those in the upper triangle of $\Ystar$, it involves the whole of $\Z$ except its major diagonal, as it requires two individual memberships for each dyad. Thirdly, each element in $\Z$ could take different roles depending on the relative positions of the nodes. In the context of citation networks, if $r$ is ahead of (behind) $s$ topologically, $\Z[rs]$ is the group membership of article $r$ when it comes to whether it is citing (cited by) article $s$. Finally, as a related point, while exactly half of $\Ystar$ contributes to the likelihood, whether a \textit{particular} dyad does so depends on the ordering $\o$. Again, if $r$ is ahead of (behind) $s$ topologically, it is $\Y[rs]$ ($\Y[sr]$) and $\C[{\Z[rs]\Z[sr]}]$ ($\C[{\Z[sr]\Z[rs]}]$) that are included in the likelihood. Given that there is no symmetry assumed on $\C$, unlike the assortative MMSBM \citep{gmgfb12, law16}, $\C[{\Z[rs]\Z[sr]}]$ and $\C[{\Z[sr]\Z[rs]}]$ are not necessarily the same, let alone the respective likelihood contributions.

\section{Likelihood and inference} \label{sect.lik_inf}
To write out the likelihood, we will mainly use the reordered versions of the matrices for two reasons. Firstly, $\Ystar$ is upper triangular, making the indexing for products easier. Secondly, the ordering $\o$ is involved in the likelihood without being written out explicitly. We just need to include an indicator variable that $\o$ is such that $\Ystar$ is upper triangular. Now, recall that, for $p<q$, $\Ystar[pq]$ is Bernoulli distributed with success probability $\C[{\Zstar[pq]\Zstar[qp]}]$. Therefore, the likelihood is 
\begin{align}
  \pi\left(\Y|\Z,\C,\o\right)=&~\one{\Ystar={\Ystar}^{+}}\times\Prod[n-1]{p=1}\Prod[n]{q=p+1}\left[\C[{\Zstar[pq]\Zstar[qp]}]^{\Ystar[pq]}\left(1-\C[{\Zstar[pq]\Zstar[qp]}]\right)^{\left(1-\Ystar[pq]\right)}\right], \nonumber
\end{align}
where $\one{E}$ is the indicator function of event $E$. The two group memberships $\Zstar[pq]$ and $\Zstar[qp]$ are already assumed to follow the multinomial distribution with probabilities $\dstar[p]$ and $\dstar[q]$, respectively, which means
\begin{align}
  \pi\left(\Z|\D,\o\right)=~\Prod[n-1]{p=1}\Prod[n]{q=p+1}\Dstar[{p\Zstar[pq]}]\Dstar[{q\Zstar[qp]}].\nonumber
\end{align}
What remains is prior specification for $\C, \D$ and $\o$ before inference can be carried out. We assume each element of $\C$ has an independent Beta prior, that is, $\C[ij]\sim\text{Beta}(\A[ij],\B[ij])$, where $\A$ and $\B$ are $K\times K$ matrices with all positive hyperparameters. We then assume that $\dstar[p]$ arises from the Dirichlet$(\a)$ distribution and is independent of $\dstar[q]$ if $p\neq q$, \textit{apriori}. This introduces the scalar parameter $\a$, which is assumed a Gamma$(3,3)$ prior (with mean $1$) here. Such choice translates to a marginal prior for $\D$ that is (approximately) uniform on the $(K-1)$-simplex. Finally, $\o$ is assumed to be uniform over all permutations of $\{1,2,\ldots,n\}$. Each of $\C$, $\a$ and $\o$ is independent of the other two parameters \textit{apriori}.

The joint posterior of $\Z,\D,\C,\a$ and $\o$, up to a proportionality constant, is given by
\begin{align}
  &\pi\left(\Z,\D,\C,\a,\o|\Y\right)\propto\pi\left(\Y,\Z,\D,\C,\a,\o\right)\nonumber\\
  =~&\pi\left(\Y|\Z,\D,\C,\a,\o\right)\times\pi\left(\Z|\D,\C,\a,\o\right)\times\pi\left(\D|\C,\a,\o\right)\times\pi\left(\C,\a,\o\right)\nonumber\\
  \propto~&\pi\left(\Y|\Z,\C,\o\right)\times\pi\left(\Z|\D,\o\right)\times\pi\left(\D|\a\right)\times\pi\left(\C\right)\times\pi\left(\a\right)\nonumber\\
  \begin{split}
  \propto~&\one{\Ystar={\Ystar}^{+}}\times\Prod[n-1]{p=1}\Prod[n]{q=p+1}\left[\C[{\Zstar[pq]\Zstar[qp]}]^{\Ystar[pq]}\left(1-\C[{\Zstar[pq]\Zstar[qp]}]\right)^{\left(1-\Ystar[pq]\right)}\right]\\
  &\times\Prod[n-1]{p=1}\Prod[n]{q=p+1}\Dstar[{p\Zstar[pq]}]\Dstar[{q\Zstar[qp]}]\times\Prod[n]{p=1}\left[\Gamma(K\a)\one{\sum_{i=1}^{K}\Dstar[pi]=1}\Prod[K]{i=1}\frac{{\Dstar[pi]}^{(\a-1)}}{\Gamma(\a)}\right]\\
  &\times\Prod[K]{i=1}\Prod[K]{j=1}\left[\C[ij]^{\A[ij]-1}\left(1-\C[ij]\right)^{\B[ij]-1}\right]\times\a^{a-1}e^{-b\a}\one{\a>0}.
  \end{split}\label{eqn.lik_inf_joint}
\end{align}

We will use Markov chain Monte Carlo (MCMC) to carry out inference, instead of the variational Bayes approach by \cite{abfx08}. The particular MCMC algorithm is simply called the \textit{regular} Gibbs sampler, in which each element of $\Z,\D$ and $\C$ is updated by a Gibbs step each while each of $\a$ and $\o$ is updated by a Metropolis step each. This regular Gibbs sampler is detailed in Appendix \ref{sect.appendix_rgs}. The stochastic gradient MCMC algorithm by \cite{law16} is not used here because of the small sample size in our application. 

As the Beta and Dirichlet distributions are indeed conditional conjugate priors for $\C$ and $\D$, respectively, it is possible to integrate out $\C$ and $\D$ to obtain $\pi(\Z,\a,\o|\Y)$, and in turn update each element of $\Z$ via a different Gibbs step without conditioning on $\C$ and $\D$. This alternative, termed the \textit{collapsed} Gibbs sampler, has been used in \cite{gs04b} for the general mixed membership model, and is potentially more efficient \textit{statistically}, that is, in terms of effective sample size per iteration for example. However, due to its complexity, there is no guarantee of higher \textit{computational} efficiency than the regular Gibbs sampler, in terms of, for example, effective sample size \textit{per unit time}. Therefore, we will stick to the regular Gibbs sampler for its implementation simplicity and computational efficiency in the application.


\section{Application} \label{sect.app}

The proposed model is fit to the citation network data presented in Section \ref{sect.eda}, with 135 nodes and 1,118 edges. With the prior knowledge of three main groups of articles detected and roughly agreed on by both algorithms in Section \ref{sect.eda}, we fit the model with $K=3,4,5$ and $6$ separately, in order to see if the articles are clustered in a similar fashion and how the clustering differs as $K$ varies. For each $K$, the regular Gibbs sampler is run to obtain a single chain of 20,000 iterations, after thinning by a factor of 500, and discarding the first 1,000,000 as burn-in. The traceplots and posterior densities of $\a$ and selected elements of $\C$, $\D$ and $\o$ are shown in figures in Appendix \ref{sect.appendix_plots}. The mixing of the chains, which appear to have converged, is reasonably good for all $K$.

The posterior densities of $\a$ for different $K$'s are overlaid in Figure \ref{fig.plot_rgs_alpha_density}. The posterior mean (and standard deviation) of $\a$ decreases with $K$, indicating that the memberships in general are getting more concentrated on fewer group(s) for each article when the number of groups increases. Also plotted is the prior density, represented by the dash-dotted line near the $x$-axis. The posterior densities for all $K$ deviate substantially from the prior density, suggesting that the model fits are not overly sensitive to the choice of prior.
\begin{knitrout}
\definecolor{shadecolor}{rgb}{0.969, 0.969, 0.969}\color{fgcolor}\begin{figure}[htbp!]

{\centering \includegraphics[width=0.49\linewidth]{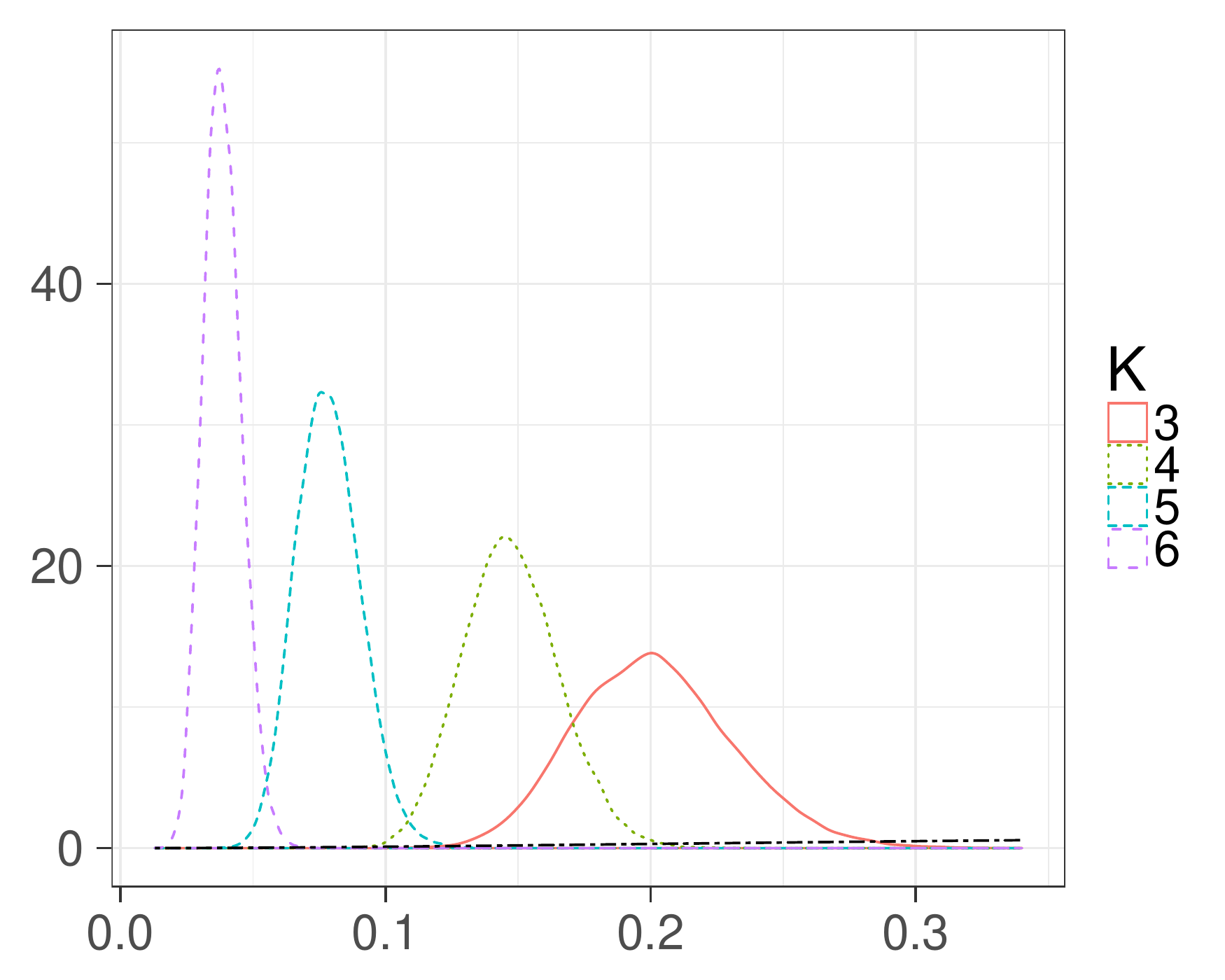} 

}

\caption[Posterior densities of $\a$ for $K$ = 3, 4, 5 and 6 overlaid]{Posterior densities of $\a$ for $K$ = 3, 4, 5 and 6 overlaid. The dashed-dotted line is the prior density.}\label{fig.plot_rgs_alpha_density}
\end{figure}

\end{knitrout}

The raster plots of the posterior means of all of $\C$ are shown in Figure \ref{fig.plot_rgs_C}. As a high probability along the major diagonal indicates a tight-knit group of articles, assuming $K=3$ results in only two main groups of articles. The remaining group, in which an article has on average a probability of less than 0.1 to cite another article in the same group, will be called the miscellaneous group. Such rule of defining miscellaneous groups applies to $K=4,5$ and $6$. The results for $K=4$ and $K=5$ seem to suggest that at least one miscellaneous group needs to be assumed in order to recover the three main groups. However, this does not necessarily imply that they match the groups mentioned in Sections \ref{sect.review} and \ref{sect.eda}. This will be checked (manually) when it comes to the memberships of the articles below. Nevertheless, such matching is conditional on a label switching of the groups, as is apparent in Figure \ref{fig.plot_rgs_C}, due to the well known identifiability issue \citep[e.g.][]{len17, mm17}.
\begin{knitrout}
\definecolor{shadecolor}{rgb}{0.969, 0.969, 0.969}\color{fgcolor}\begin{figure}[htbp!]

{\centering \includegraphics[width=0.49\linewidth]{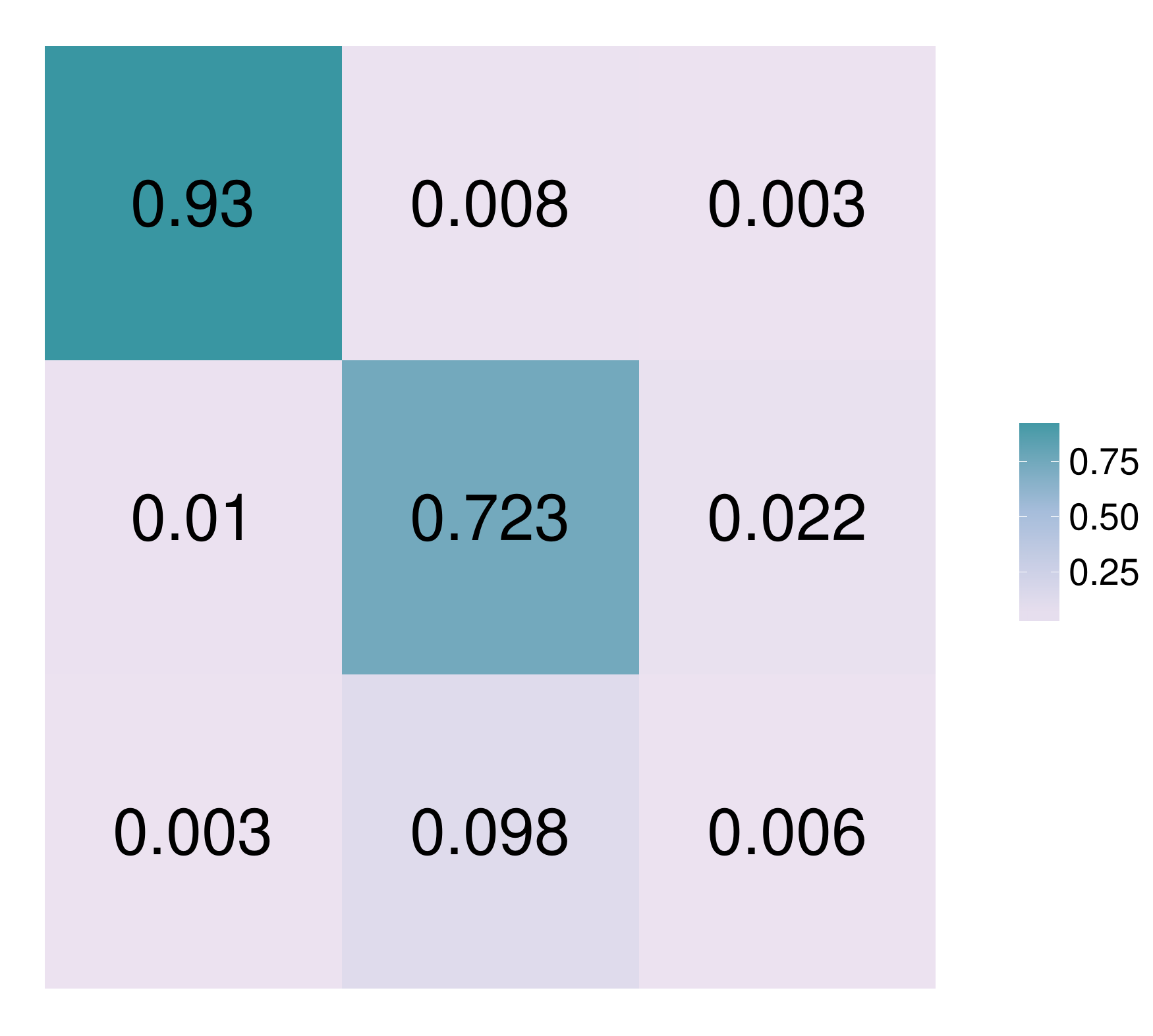} 
\includegraphics[width=0.49\linewidth]{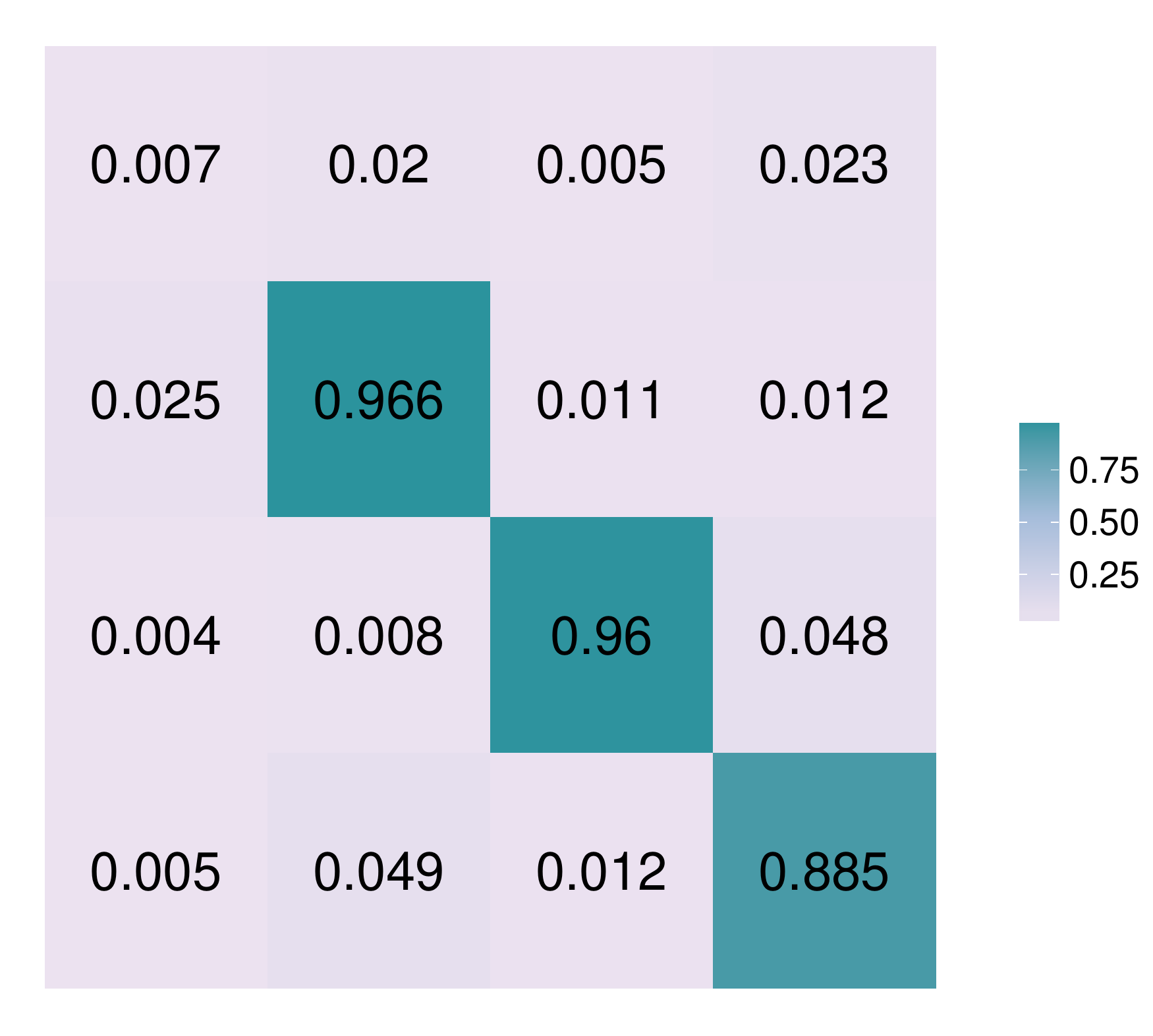} 
\includegraphics[width=0.49\linewidth]{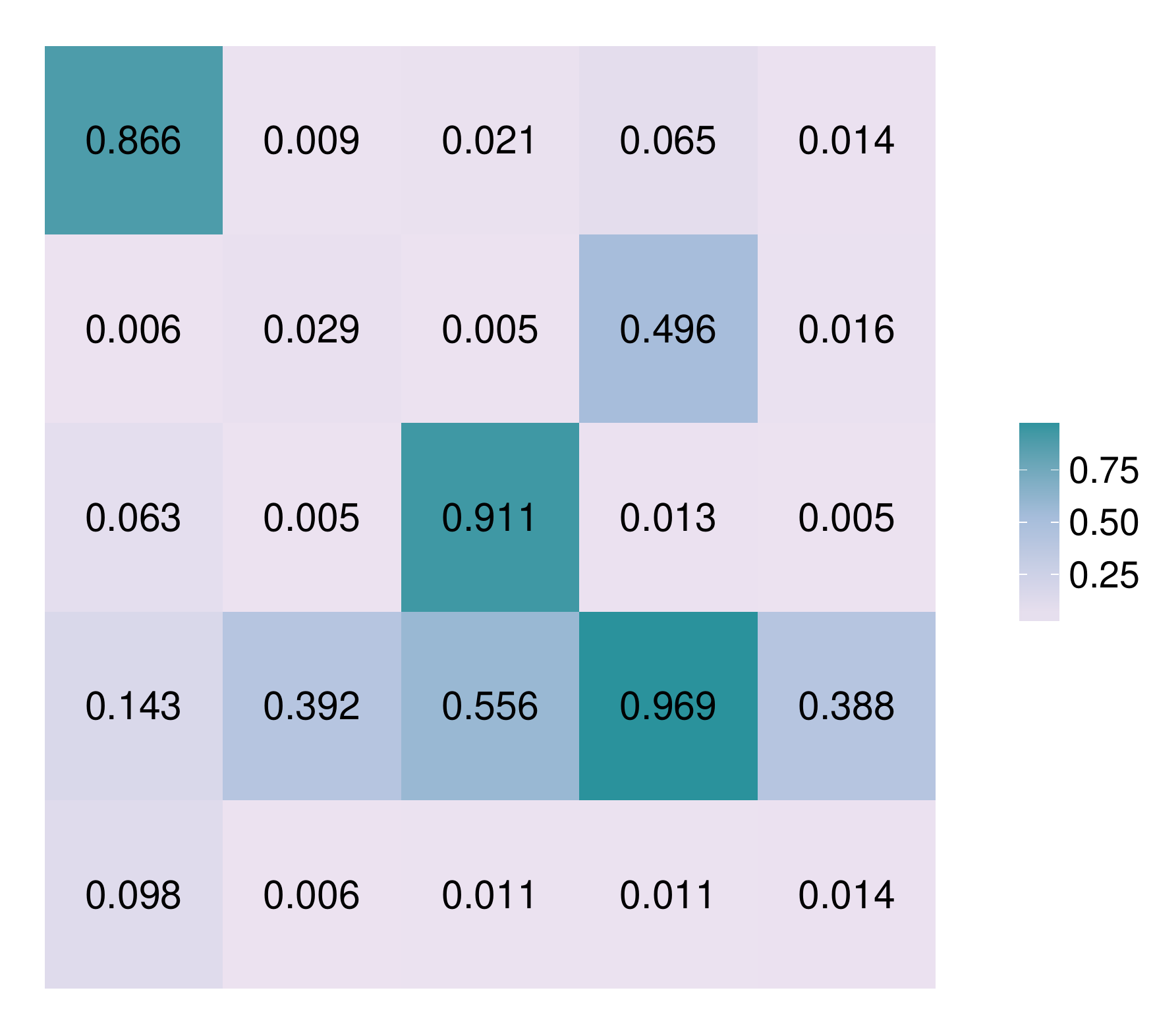} 
\includegraphics[width=0.49\linewidth]{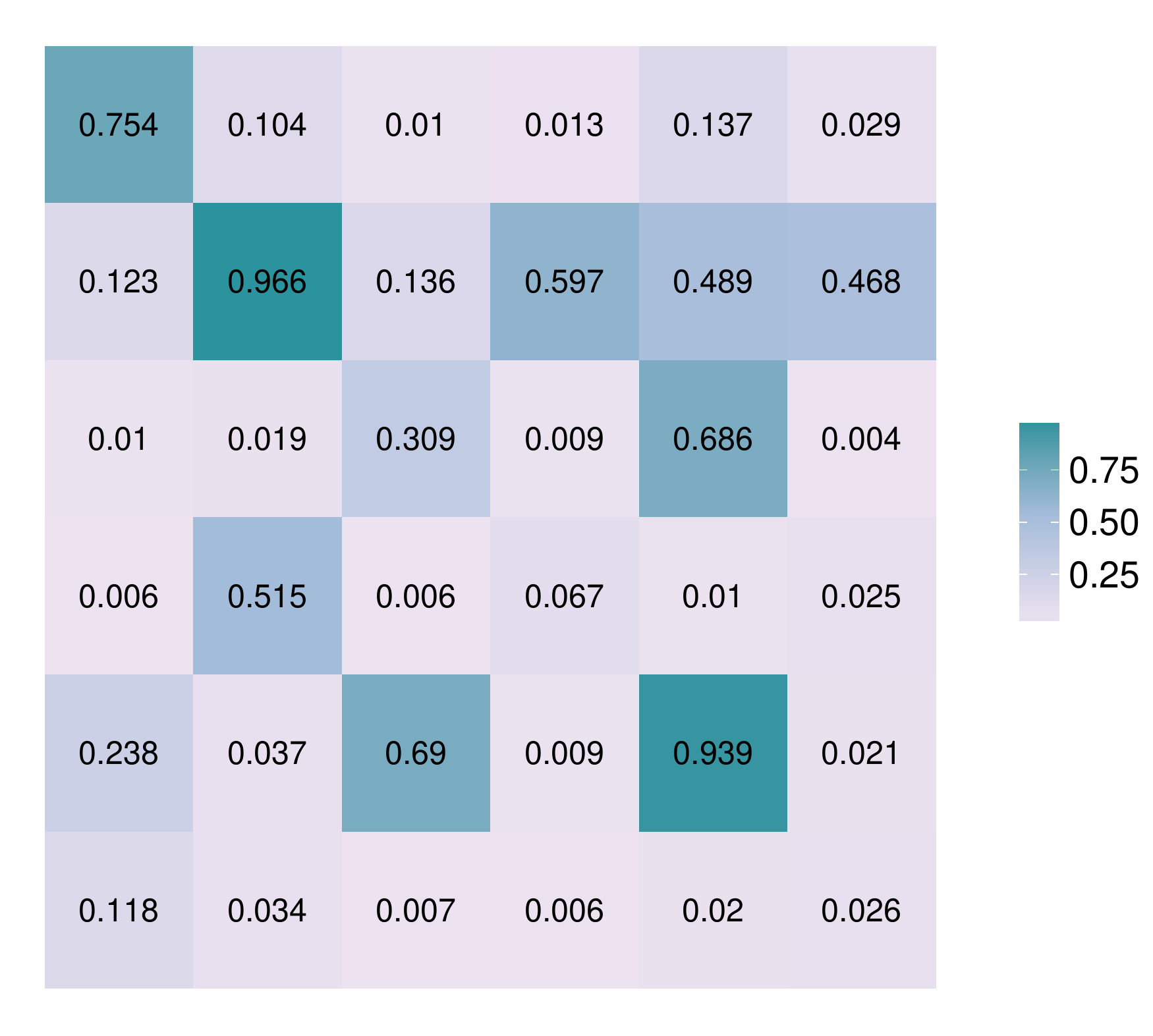} 

}

\caption[Raster plot of posterior mean of $\C$ for $K$ = 3 (top left), 4 (top right), 5 (bottom left) and 6 (bottom right)]{Raster plot of posterior mean of $\C$ for $K$ = 3 (top left), 4 (top right), 5 (bottom left) and 6 (bottom right).}\label{fig.plot_rgs_C}
\end{figure}

\end{knitrout}

The heatmap of the memberships of the articles is plotted in Figure \ref{fig.plot_rgs_D_heatmap}, with the clustering simply based on the model results. Specifically, if article A has a membership probability greater than $1/K$ in at least one of the main groups, then article A is put into the main group with the highest membership probability. Otherwise, it is put into the miscellaneous group with the highest membership probability. Note that such preference of main groups to miscellaneous groups is only for the sake of visualisation. The closer the colours along one row, the more mixed membership the corresponding article has.
\begin{knitrout}
\definecolor{shadecolor}{rgb}{0.969, 0.969, 0.969}\color{fgcolor}\begin{figure}[htbp!]

{\centering \includegraphics[width=0.49\linewidth]{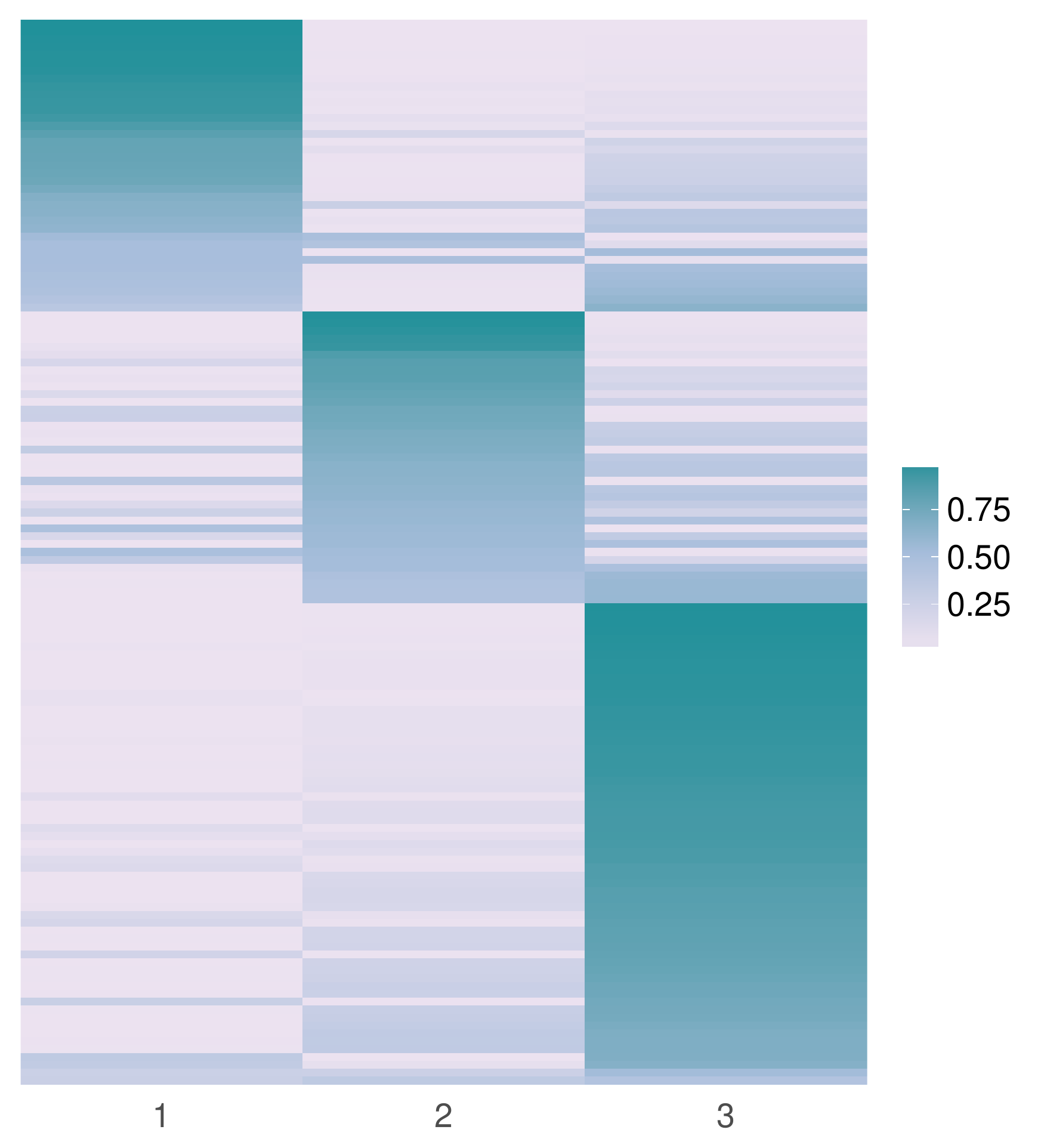} 
\includegraphics[width=0.49\linewidth]{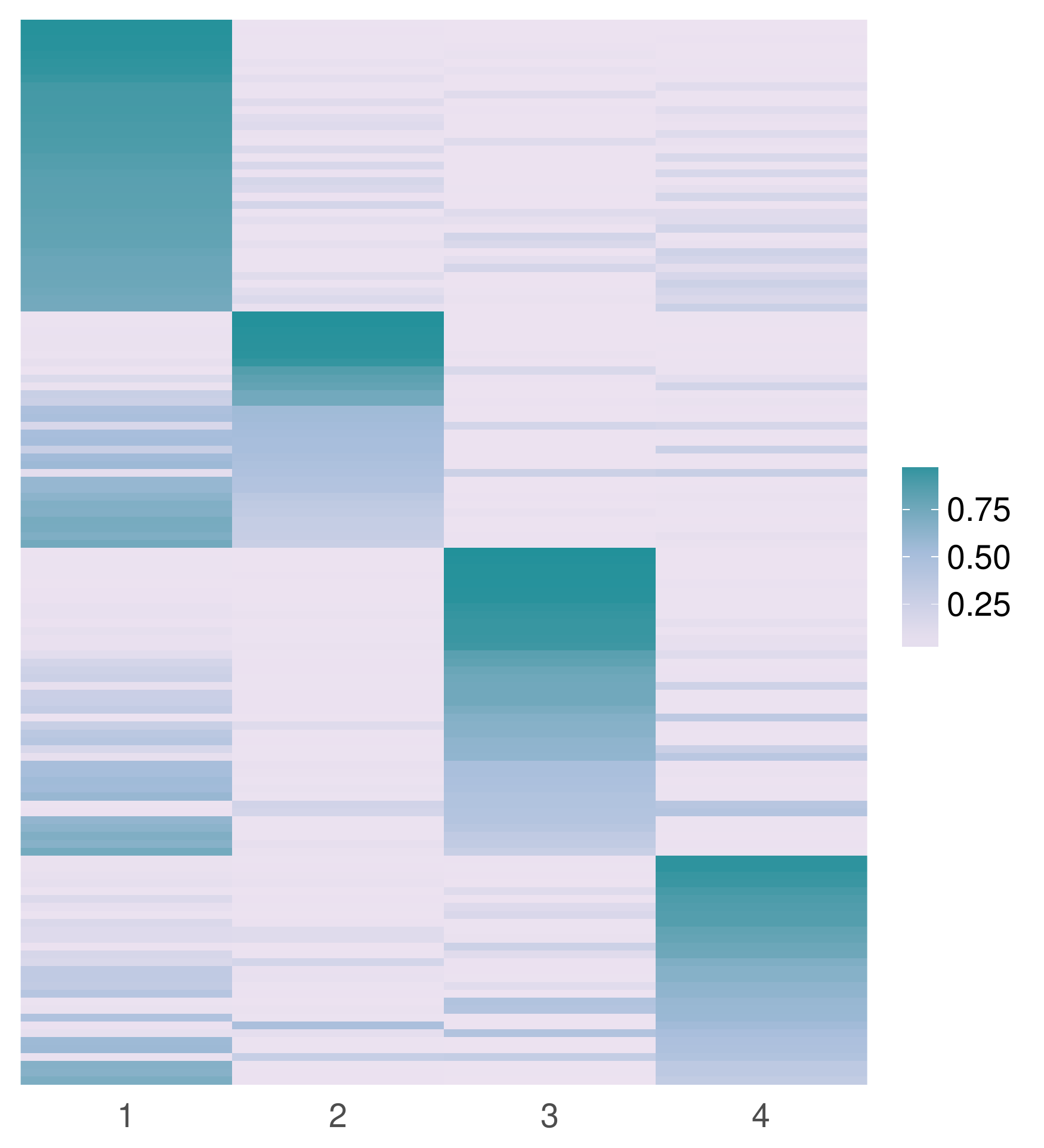} 
\includegraphics[width=0.49\linewidth]{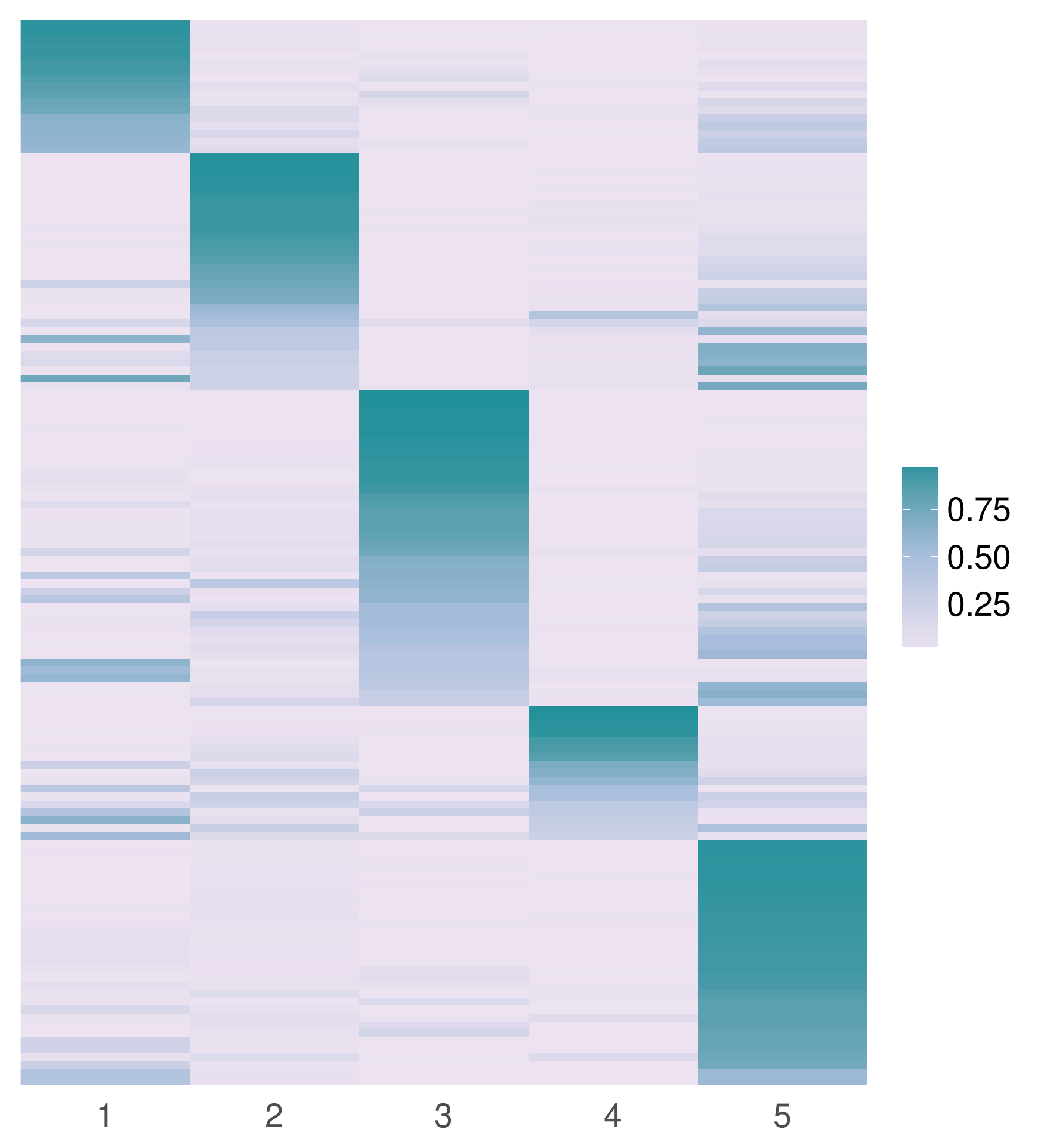} 
\includegraphics[width=0.49\linewidth]{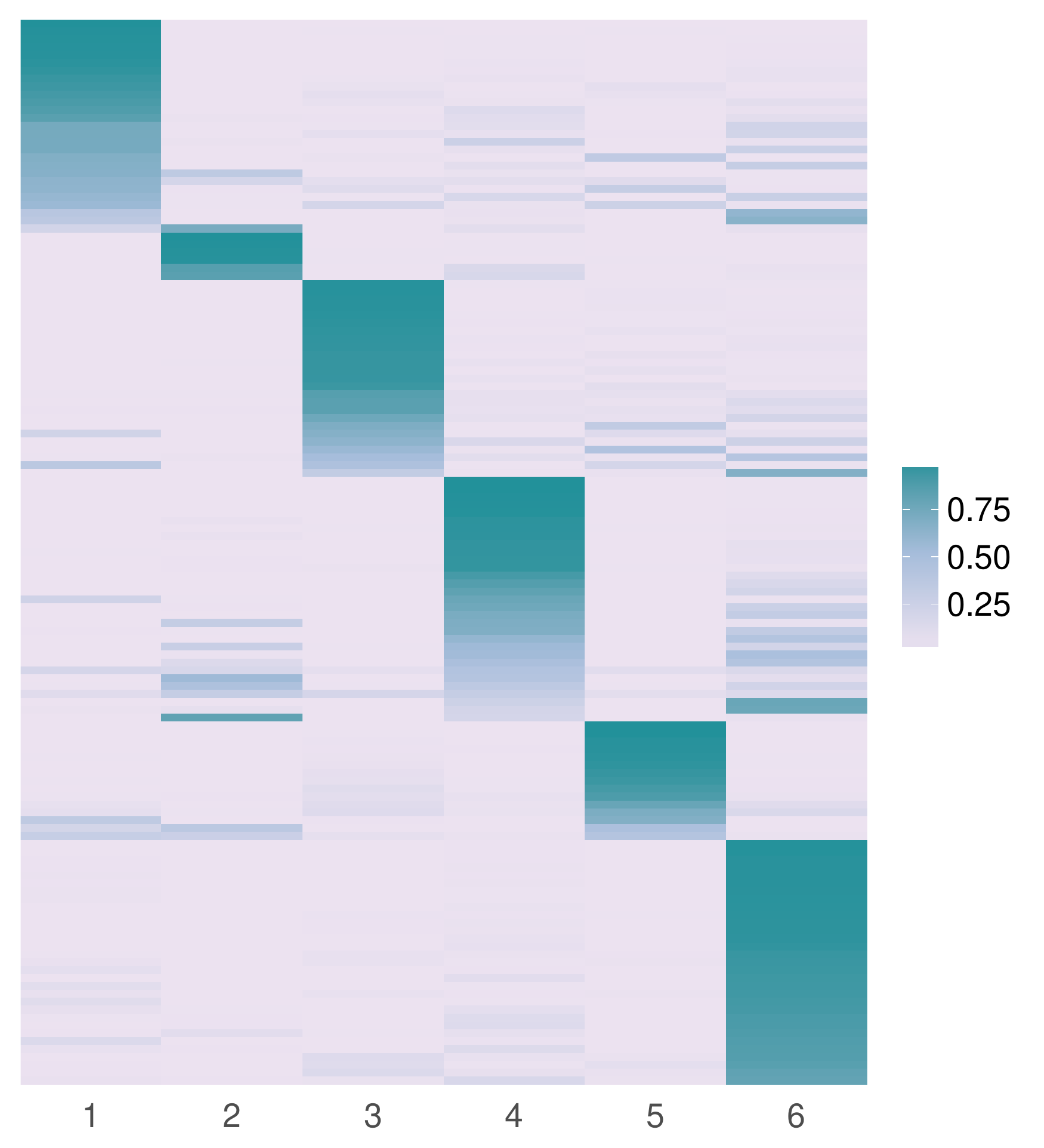} 

}

\caption[Heatmap of the memberships of the articles for $K$ = 3 (top left), 4 (top right), 5 (bottom left) and 6 (bottom right)]{Heatmap of the memberships of the articles for $K$ = 3 (top left), 4 (top right), 5 (bottom left) and 6 (bottom right).}\label{fig.plot_rgs_D_heatmap}
\end{figure}

\end{knitrout}

Another way of visualising the memberships is the network plot similar to Figure \ref{fig.plot_network_subgraph}, but with some aspects of the individual memberships shown. Shown in Figure \ref{fig.plot_rgs_D_network_subgraph_K_4} is the modified network plot for $K=4$, where the colours represent the highest (main group) memberships defined for Figure \ref{fig.plot_rgs_D_heatmap}, and the sizes represent the respective posterior means of membership probabilities. This means, in general, the smaller the node is, the more mixed the memberships of the article is. It is clear that the colours are coherent with the clustering of the articles. Furthermore, upon manual checking, the three main groups match well with those in Sections \ref{sect.review} and \ref{sect.eda}, suggesting that detection of the 3 main groups is fairly robust to the specific choice of model adopted. More importantly, the model results provide the mixed memberships of the articles, which are not possible by the hard clustering of our review or the community detecting algorithms. The corresponding plot for $K=5$ is in Figure \ref{fig.plot_rgs_D_network_subgraph_K_5}. Here, assuming one additional latent group results in one original main group (when $K=4$) being split into two partitions, with the more central articles forming one smaller group (when $K=5$). The more peripheral ones merge with some miscellaneous articles (when $K=4$) to form one ``new'' group (when $K=5$). Such changes in the clustering explains why some off-diagonal group-to-group probabilities soar, as illustrated in Figure \ref{fig.plot_rgs_C}, when $K$ increase from 4 to 5.
\begin{knitrout}
\definecolor{shadecolor}{rgb}{0.969, 0.969, 0.969}\color{fgcolor}\begin{figure}[htbp!]

{\centering \includegraphics[width=0.65\linewidth]{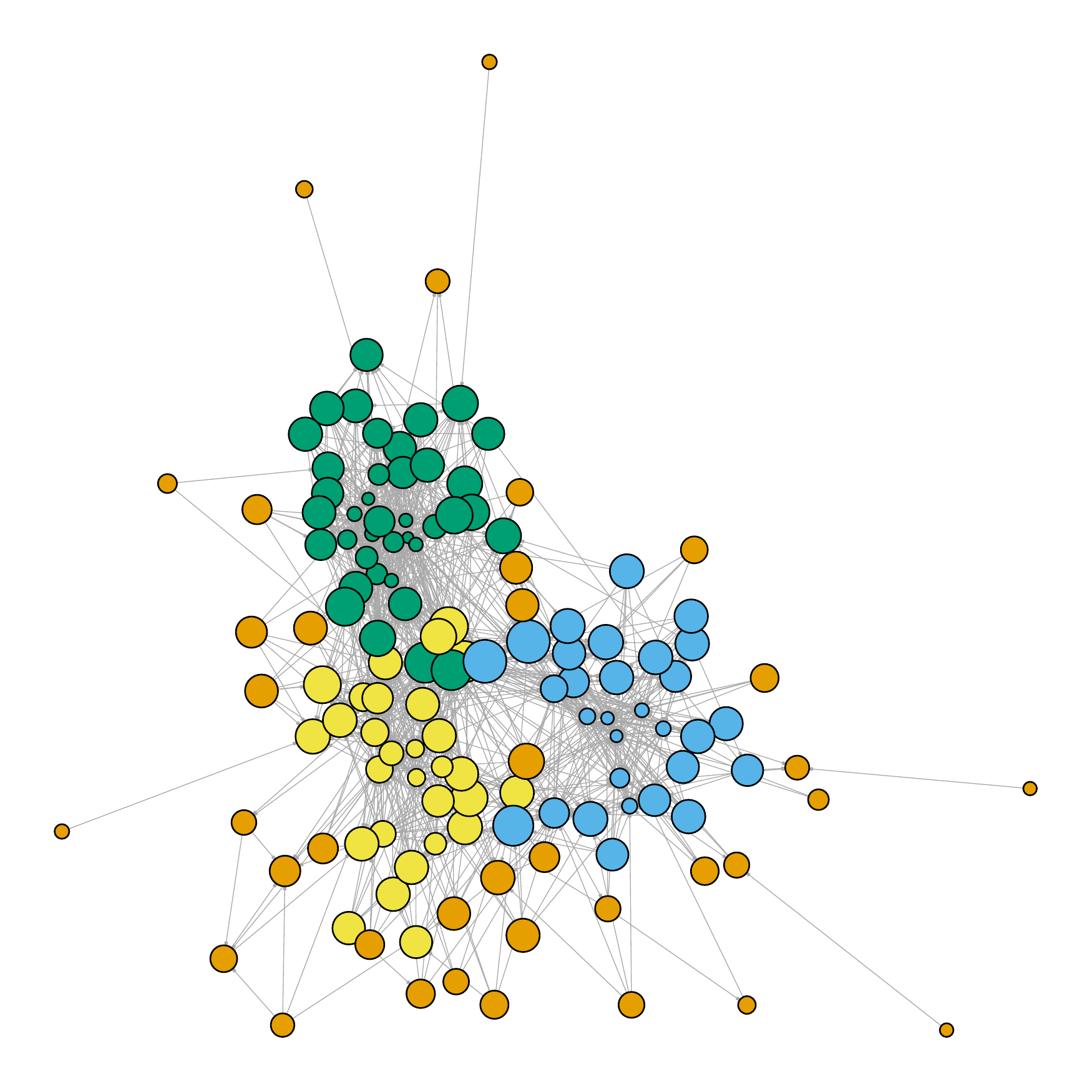} 

}

\caption[Network plot of citing articles, with colours representing the highest (main group) memberships and sizes representing the respective membership probabilites, for $K=4$]{Network plot of citing articles, with colours representing the highest (main group) memberships and sizes representing the respective membership probabilites, for $K=4$.}\label{fig.plot_rgs_D_network_subgraph_K_4}
\end{figure}

\end{knitrout}

\begin{knitrout}
\definecolor{shadecolor}{rgb}{0.969, 0.969, 0.969}\color{fgcolor}\begin{figure}[htbp!]

{\centering \includegraphics[width=0.65\linewidth]{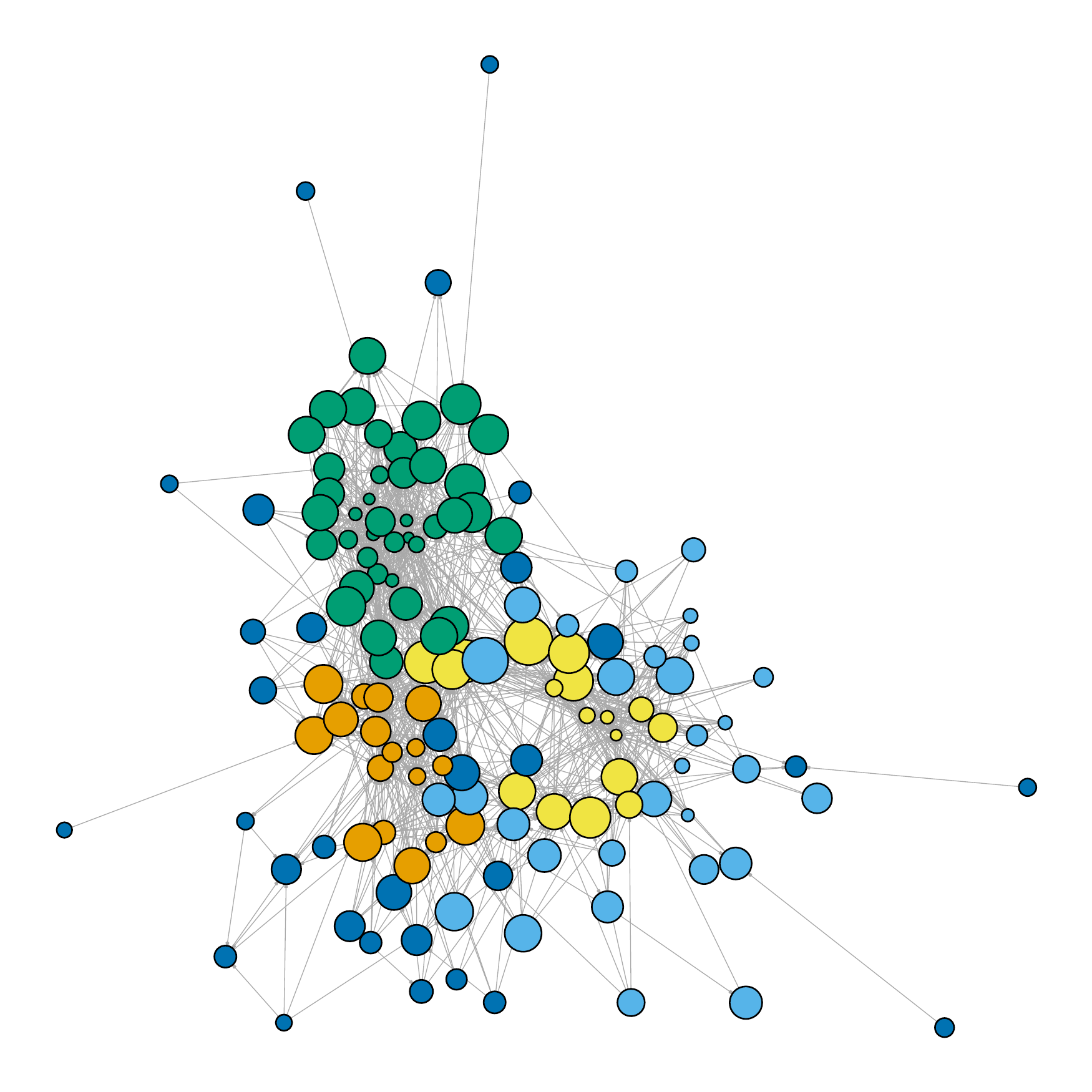} 

}

\caption[Network plot of citing articles, with colours representing the highest (main group) memberships and sizes representing the respective membership probabilites, for $K=5$]{Network plot of citing articles, with colours representing the highest (main group) memberships and sizes representing the respective membership probabilites, for $K=5$.}\label{fig.plot_rgs_D_network_subgraph_K_5}
\end{figure}

\end{knitrout}


One last visualisation regarding the group memberships is the projection of their posterior means to the $(K-1)$-simplex, similar to what \cite{abfx08} did for the 18 monks in their analysis. We present such projection for $K=4$ to the 3-simplex here, which is a regular tetrahedron, in Figure \ref{fig.plot_rgs_proj_K_4}. The 12 reviews are highlighed by the first letters of the first authors, while the remaining 123 articles are plotted as dots. Shown is the ``main'' surface of which the three vertices represent the three main groups. The size of the point represents how close the corresponding article is to this main surface, and an article having 100\% membership in the miscellaneous group is represented by a dot of size 0 at the centre of the equilateral triangle. Most of the articles lie along the edges between the main groups and the miscellaneous group, meaning that a typical article belongs largely, but to a varying degree, to only one main group. All the reviews have a similar and close to largest possible size to the main surface, indicating a negligible membership in the miscellaneous group, with \cite{ke05} as the only exception. This is because the said review concerns epidemics on networks, which is not a major topic in our data. The memberships of the other reviews clearly show their different positions in the literature. Some of them have almost entire memberships in only one of the main groups, namely \cite{ab02a} and \cite{newman03a} for the generative models, \cite{hks12} for the ERGMs, and \cite{mr14} for the latent models. Some of them have mixed memberships between two main groups, namely \cite{oo14} and \cite{snijders11} between the ERGMs and the latent models, and \cite{fortunato10} between the generative models and the latent models. Finally, the rest of the reviews, in particular \cite{channarond15}, have mixed memberships between all three main groups, suggesting that they have cited and/or been cited by a wide spectrum of articles in the literature.
\begin{knitrout}
\definecolor{shadecolor}{rgb}{0.969, 0.969, 0.969}\color{fgcolor}\begin{figure}[htbp!]

{\centering \includegraphics[width=0.95\linewidth]{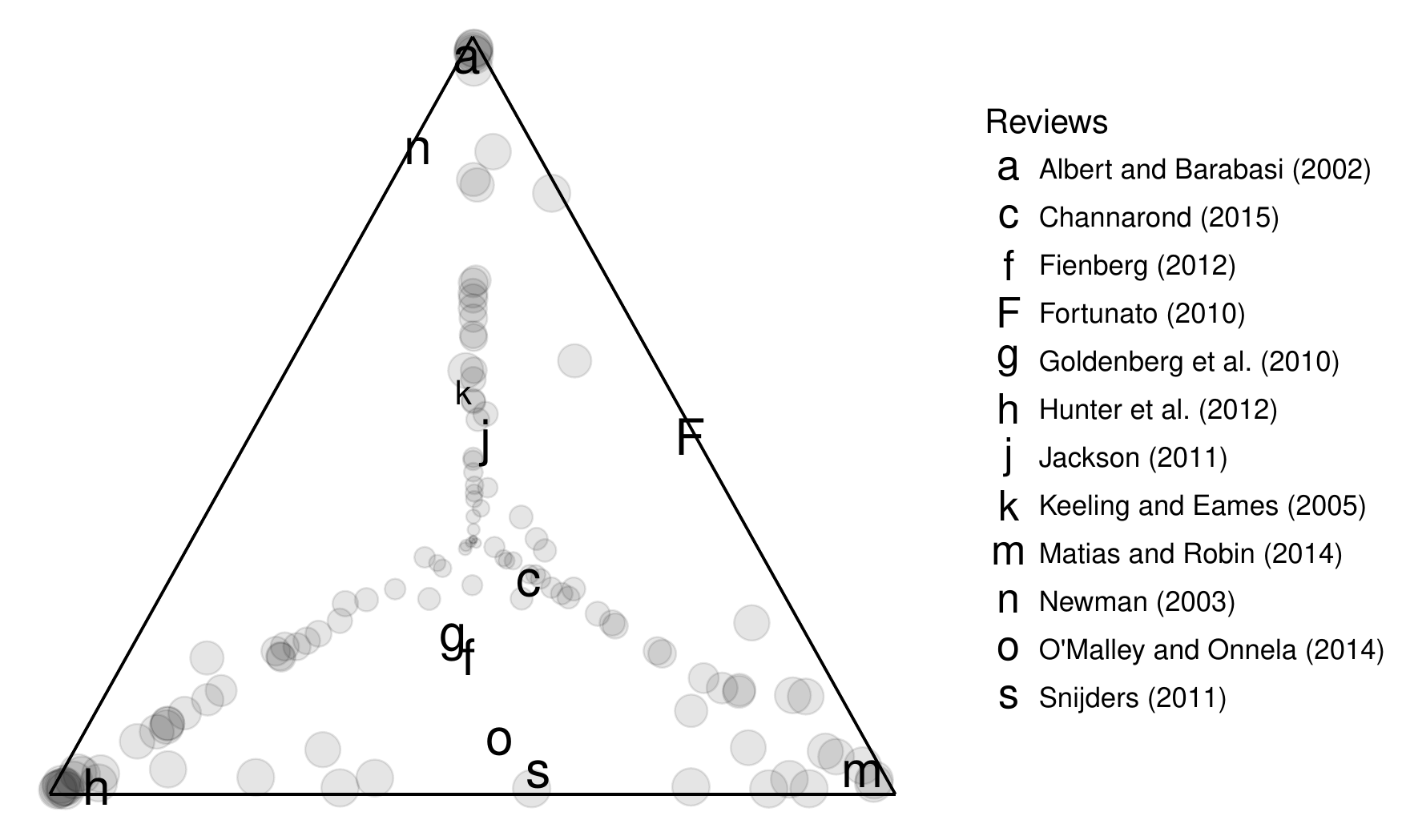} 

}

\caption[Projection of posterior means of membership probabilities to the 3-simplex for $K=4$]{Projection of posterior means of membership probabilities to the 3-simplex for $K=4$.}\label{fig.plot_rgs_proj_K_4}
\end{figure}

\end{knitrout}

To dissect the posterior of the topological order, we convert $\o$ to the positions of the articles in it, and look at the posterior of these individual positions. Their posterior densities for different $K$ are overlaid in Figure \ref{fig.plot_rgs_i_density}. In general, the positions, and in turn $\o$ as a whole, are not dependent on the choice of $K$. Alternatively, these positions can be visualised by a ``heatmap'' in Figure \ref{fig.plot_rgs_i_heatmap}, where the colour across each row represents the posterior density of the position of an article. A wide horizontal bar means a large variation in the posterior, indicating that the article could take up a wide range of positions, possible because it is not citing many other articles in the data and/or is not cited frequently.
\begin{knitrout}
\definecolor{shadecolor}{rgb}{0.969, 0.969, 0.969}\color{fgcolor}\begin{figure}[htbp!]

{\centering \includegraphics[width=0.95\linewidth]{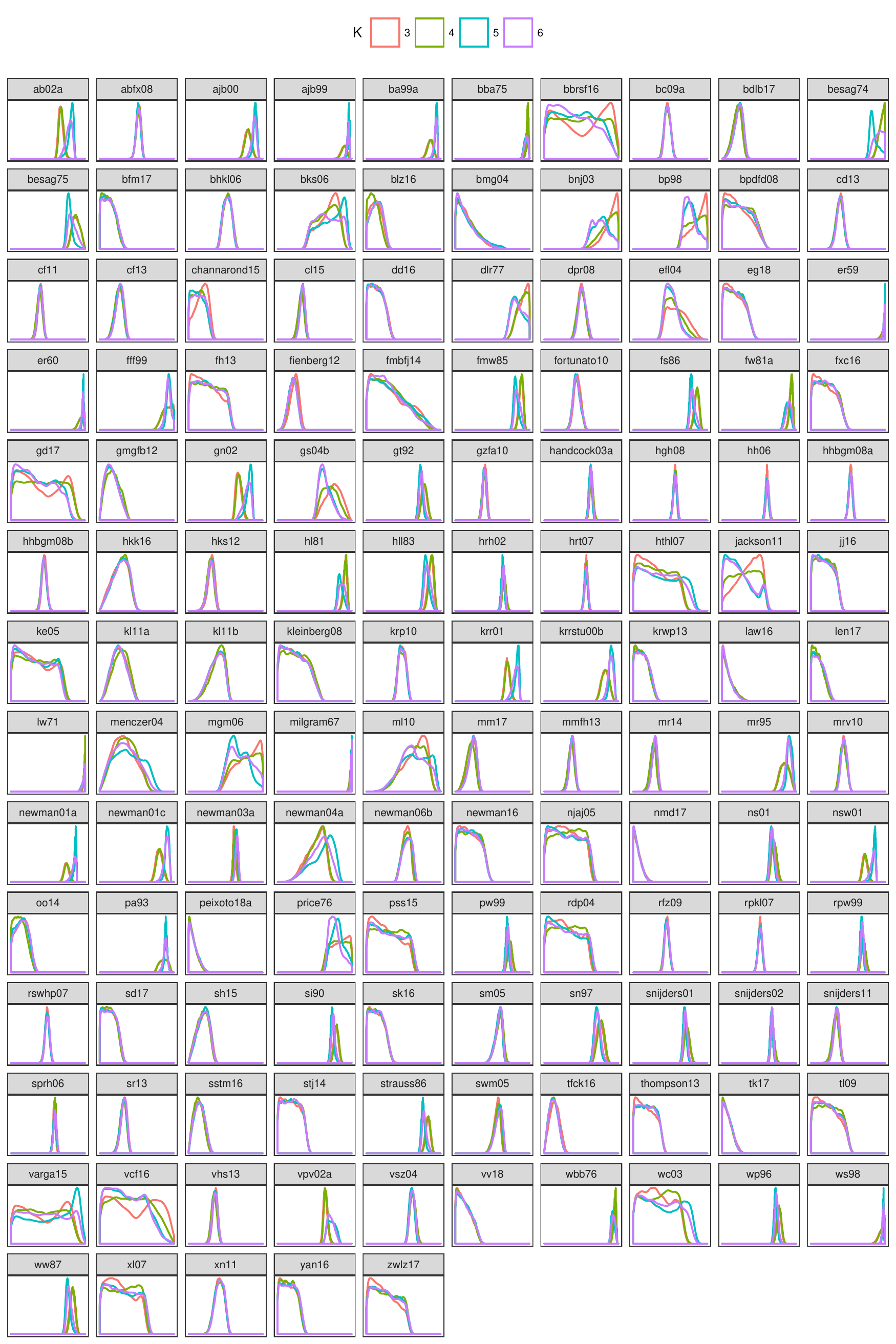} 

}

\caption[Posterior densities of positions in topological order for $K$ = 3, 4, 5 and 6 overlaid]{Posterior densities of positions in topological order for $K$ = 3, 4, 5 and 6 overlaid.}\label{fig.plot_rgs_i_density}
\end{figure}

\end{knitrout}

\begin{knitrout}
\definecolor{shadecolor}{rgb}{0.969, 0.969, 0.969}\color{fgcolor}\begin{figure}[htbp!]

{\centering \includegraphics[width=0.49\linewidth]{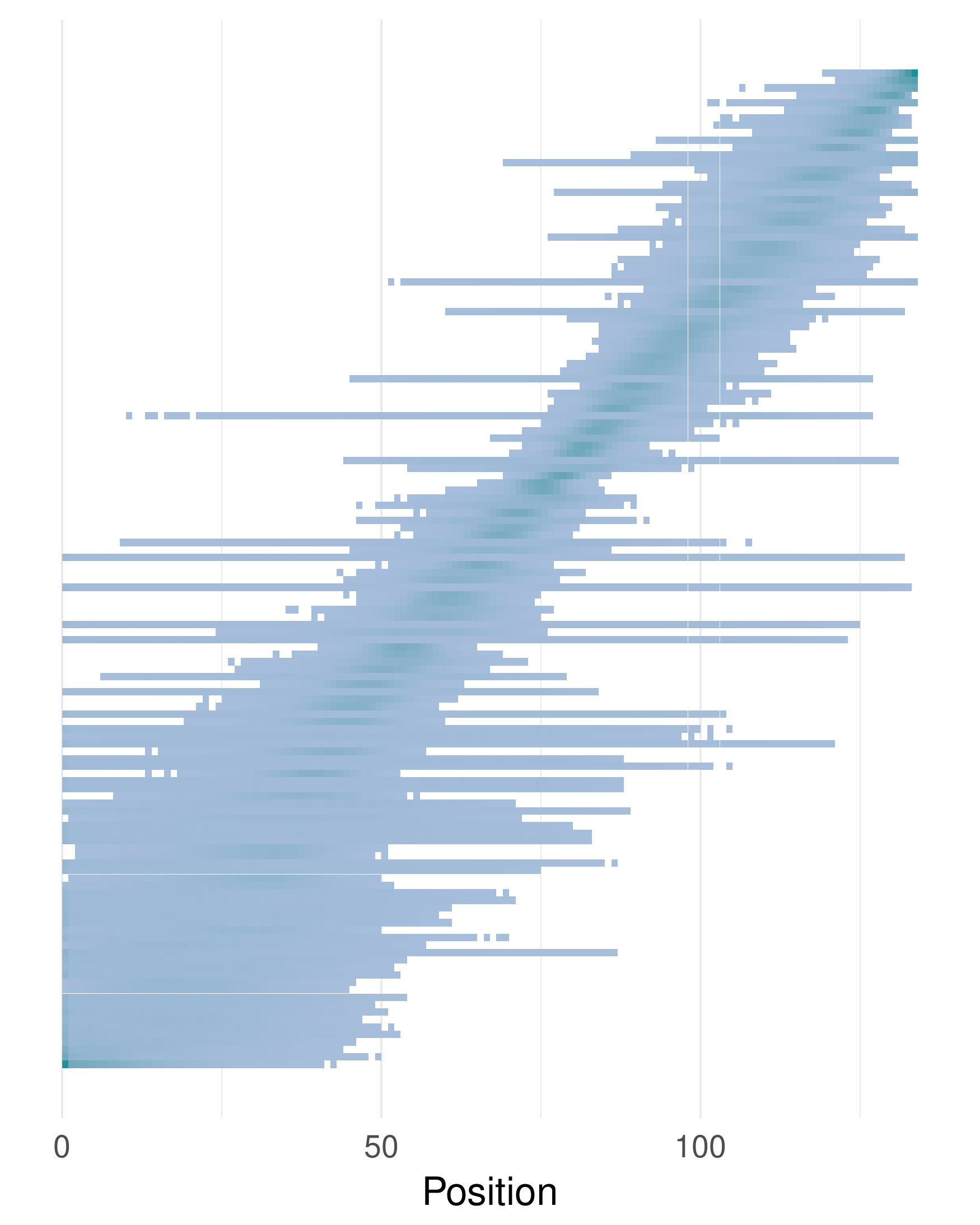} 
\includegraphics[width=0.49\linewidth]{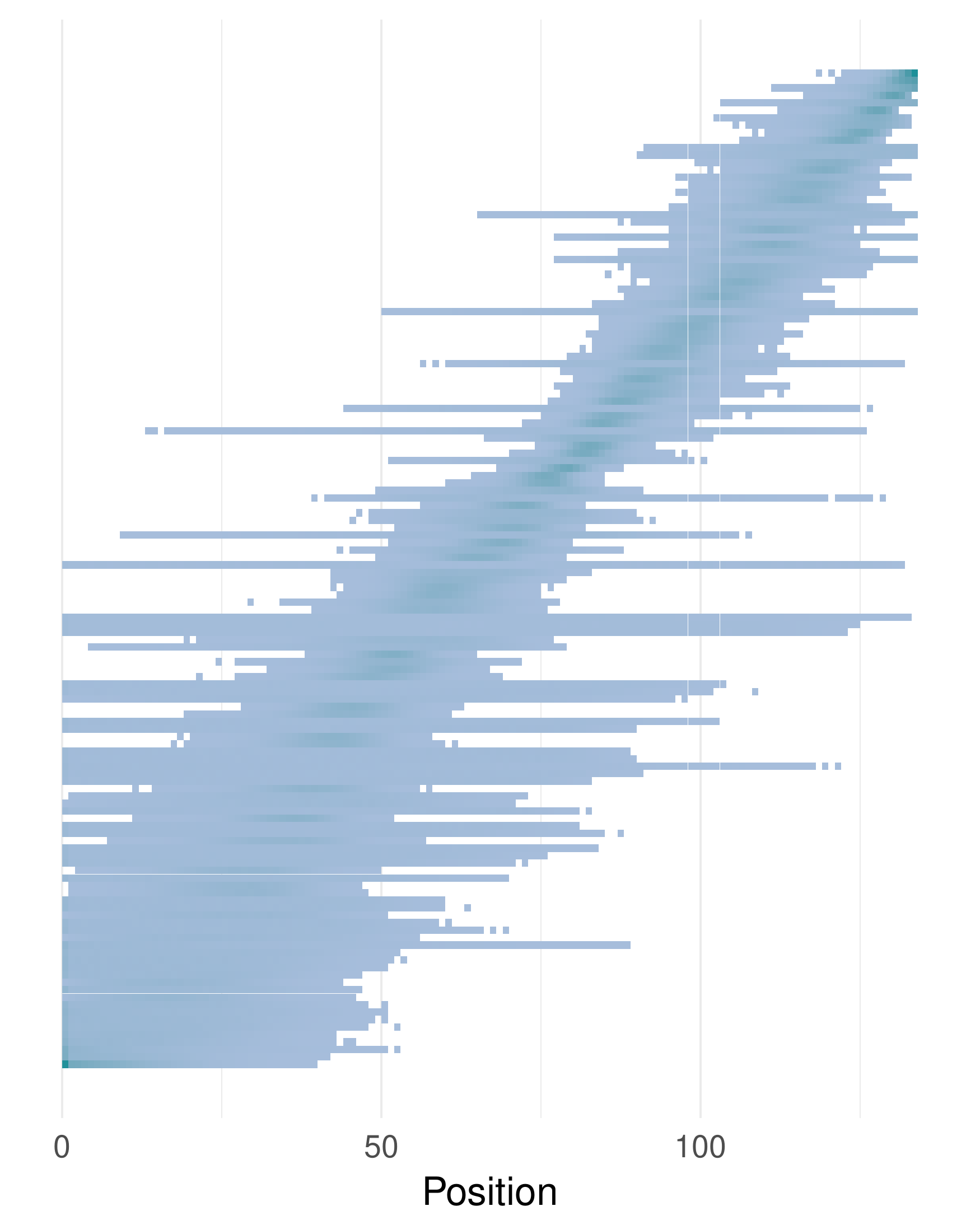} 
\includegraphics[width=0.49\linewidth]{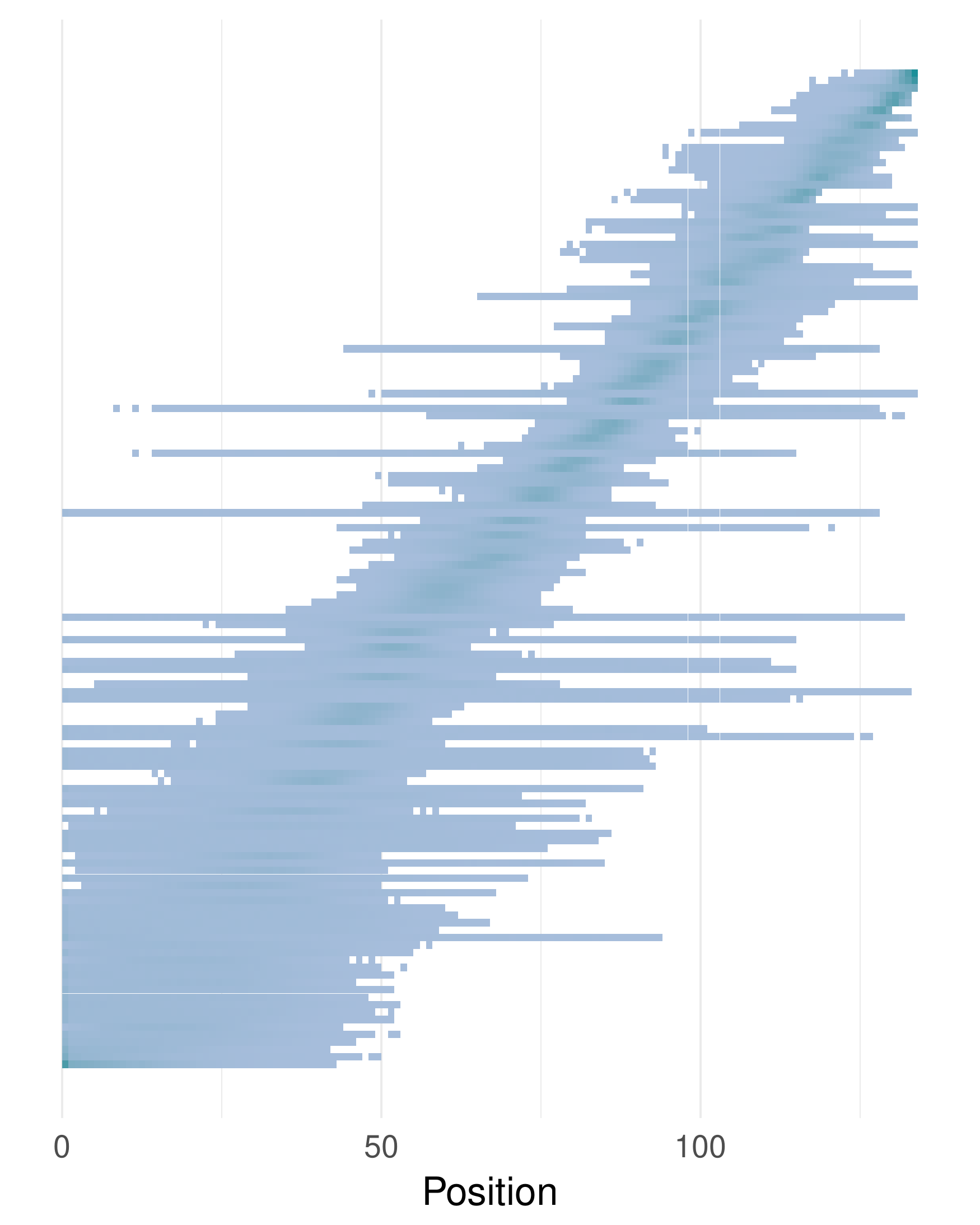} 
\includegraphics[width=0.49\linewidth]{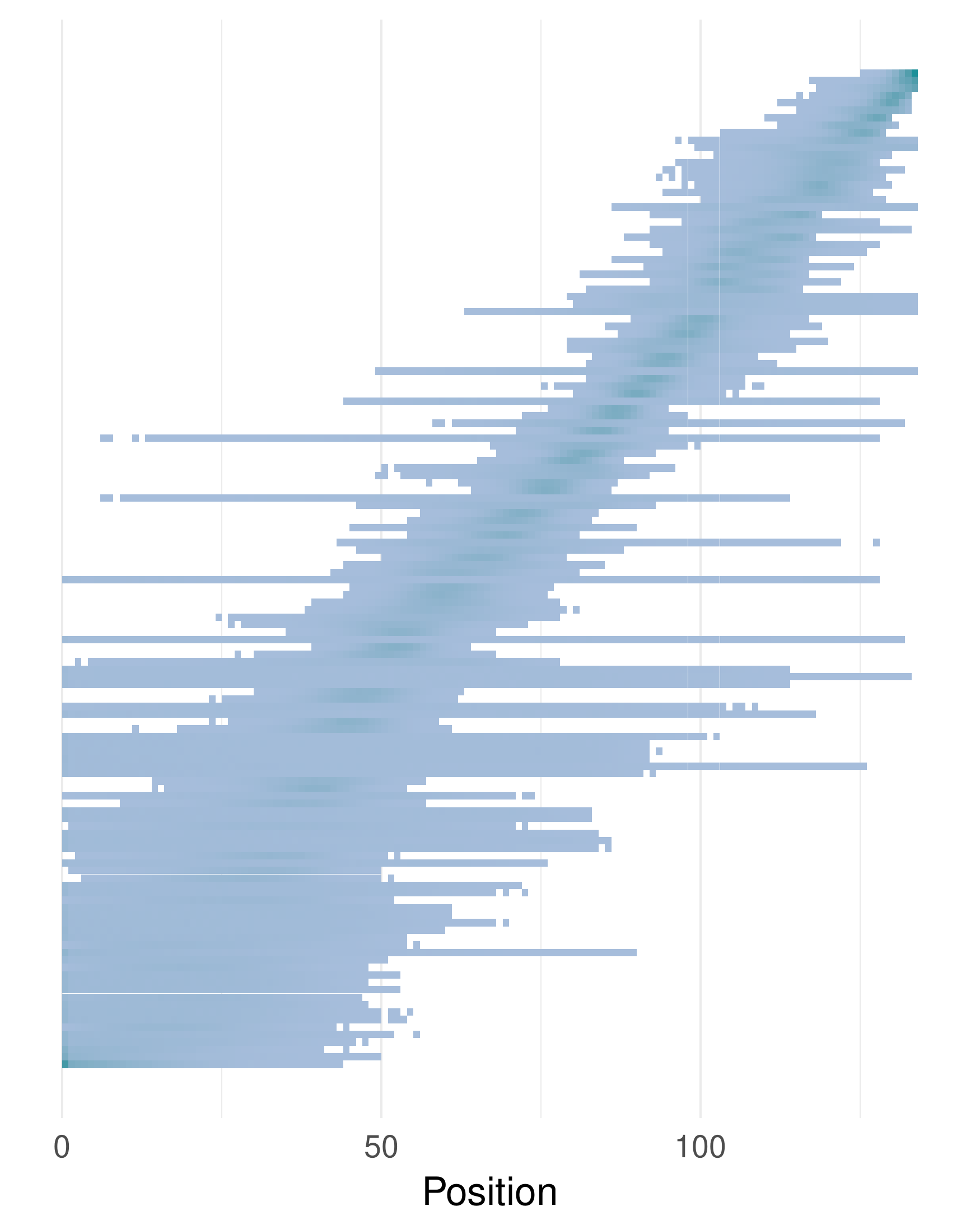} 

}

\caption[Heatmap of the posterior distributions of the positions in $\o$ for $K$ = 3 (top left), 4 (top right), 5 (bottom left) and 6 (bottom right)]{Heatmap of the posterior distributions of the positions in $\o$ for $K$ = 3 (top left), 4 (top right), 5 (bottom left) and 6 (bottom right). The articles are sorted in descending order of the posterior mean of the position}\label{fig.plot_rgs_i_heatmap}
\end{figure}

\end{knitrout}

Recall that, if article A cites article B, then A is topologically ahead of B. However, this also means that, in most cases, article A has a later year of publication than article B. In some sense, the topological order should be similar to the reverse chronological order. Therefore, the mean positions are plotted against the years of publication in Figure \ref{fig.plot_rgs_i_vs_year}. While the negative correlation is expected, the relationship between the two types of ordering is not quite linear.
\begin{knitrout}
\definecolor{shadecolor}{rgb}{0.969, 0.969, 0.969}\color{fgcolor}\begin{figure}[htbp!]

{\centering \includegraphics[width=0.49\linewidth]{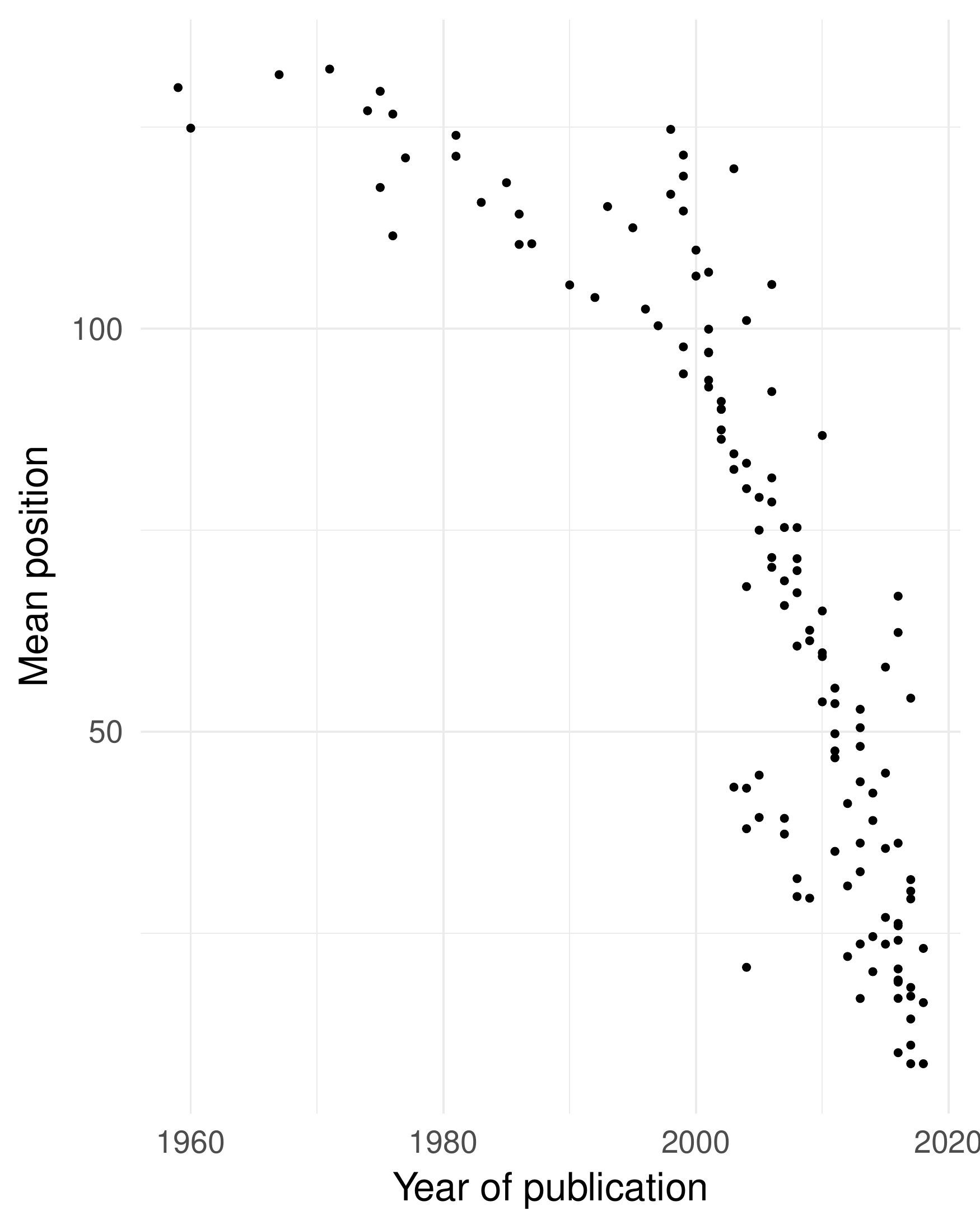} 
\includegraphics[width=0.49\linewidth]{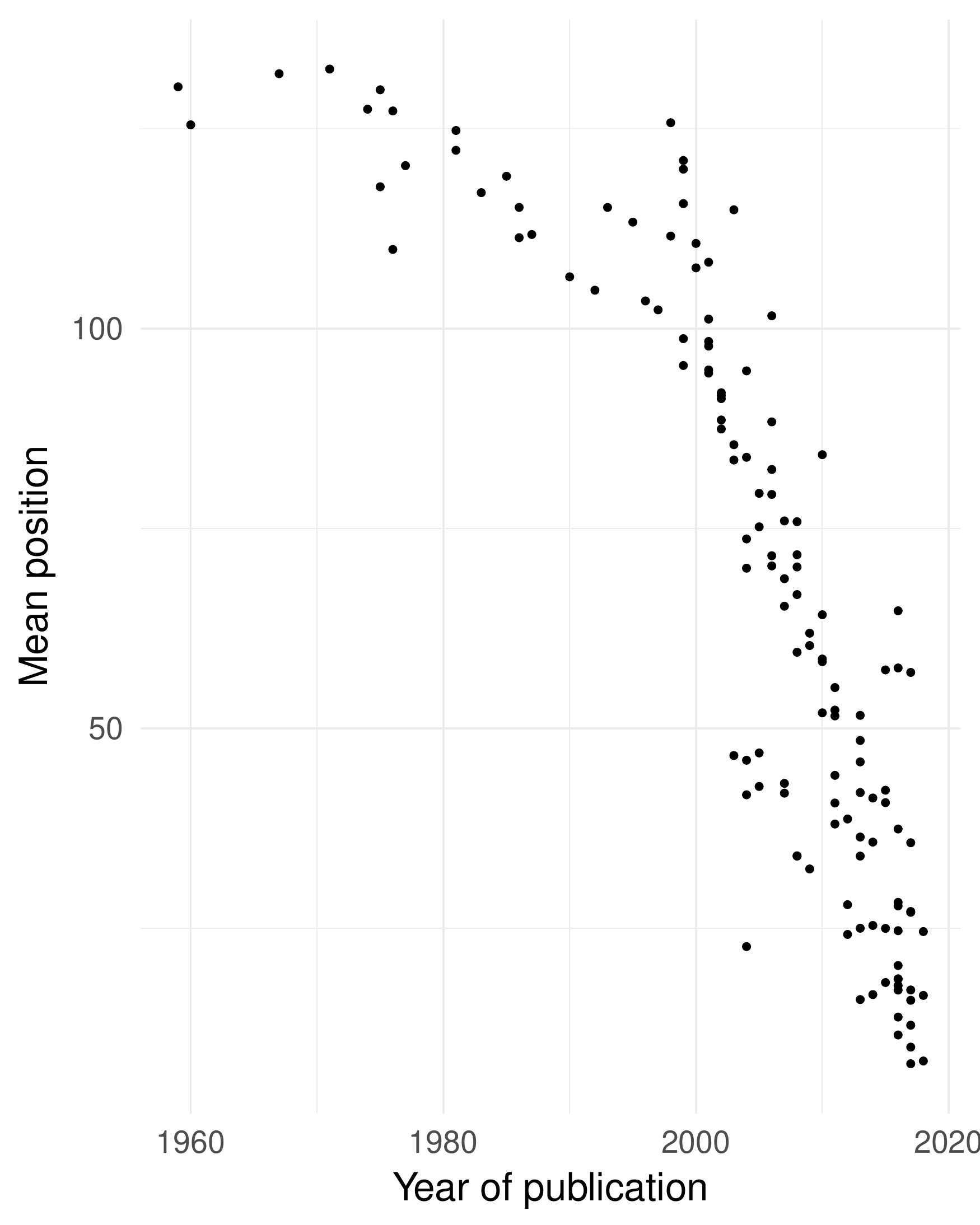} 
\includegraphics[width=0.49\linewidth]{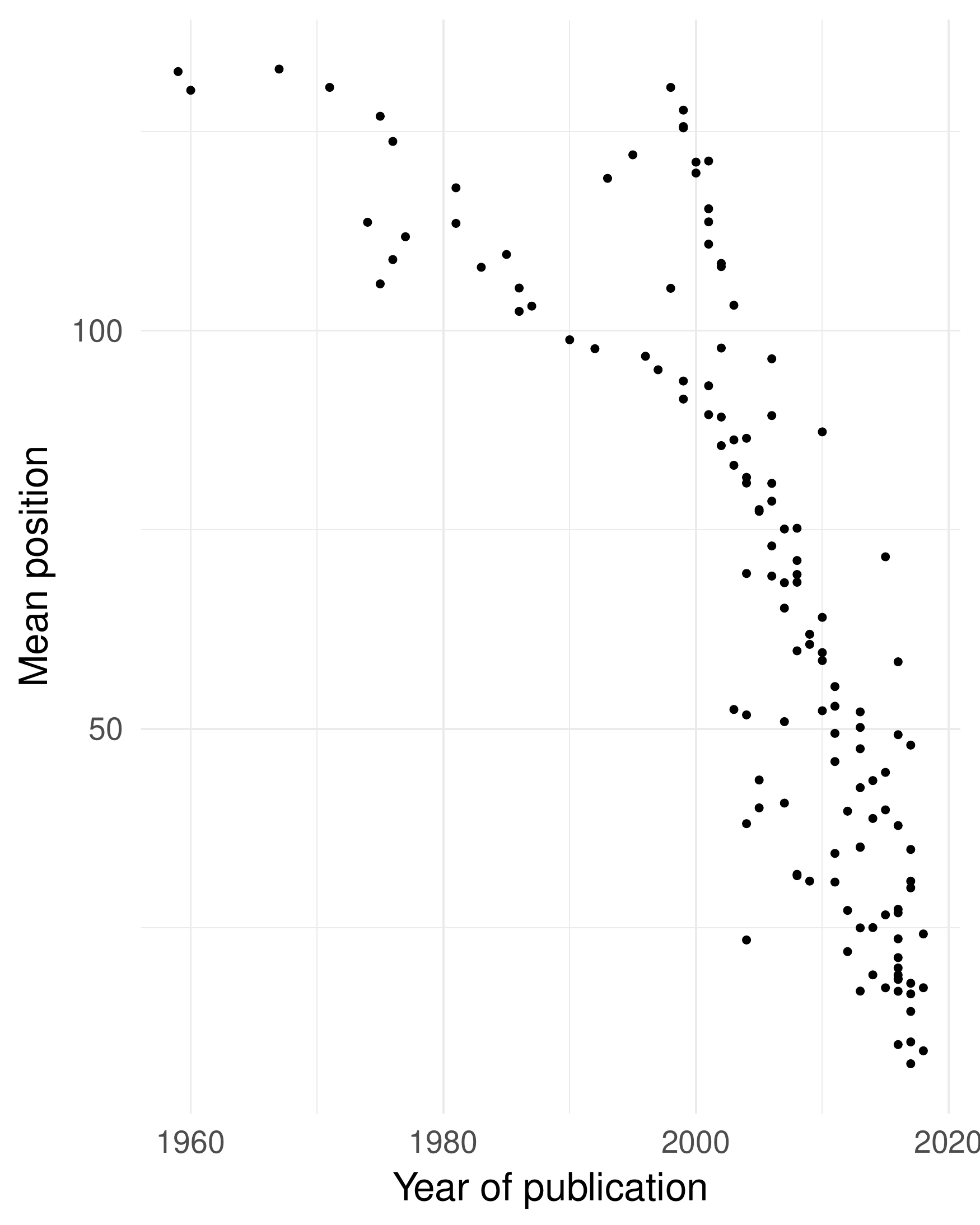} 
\includegraphics[width=0.49\linewidth]{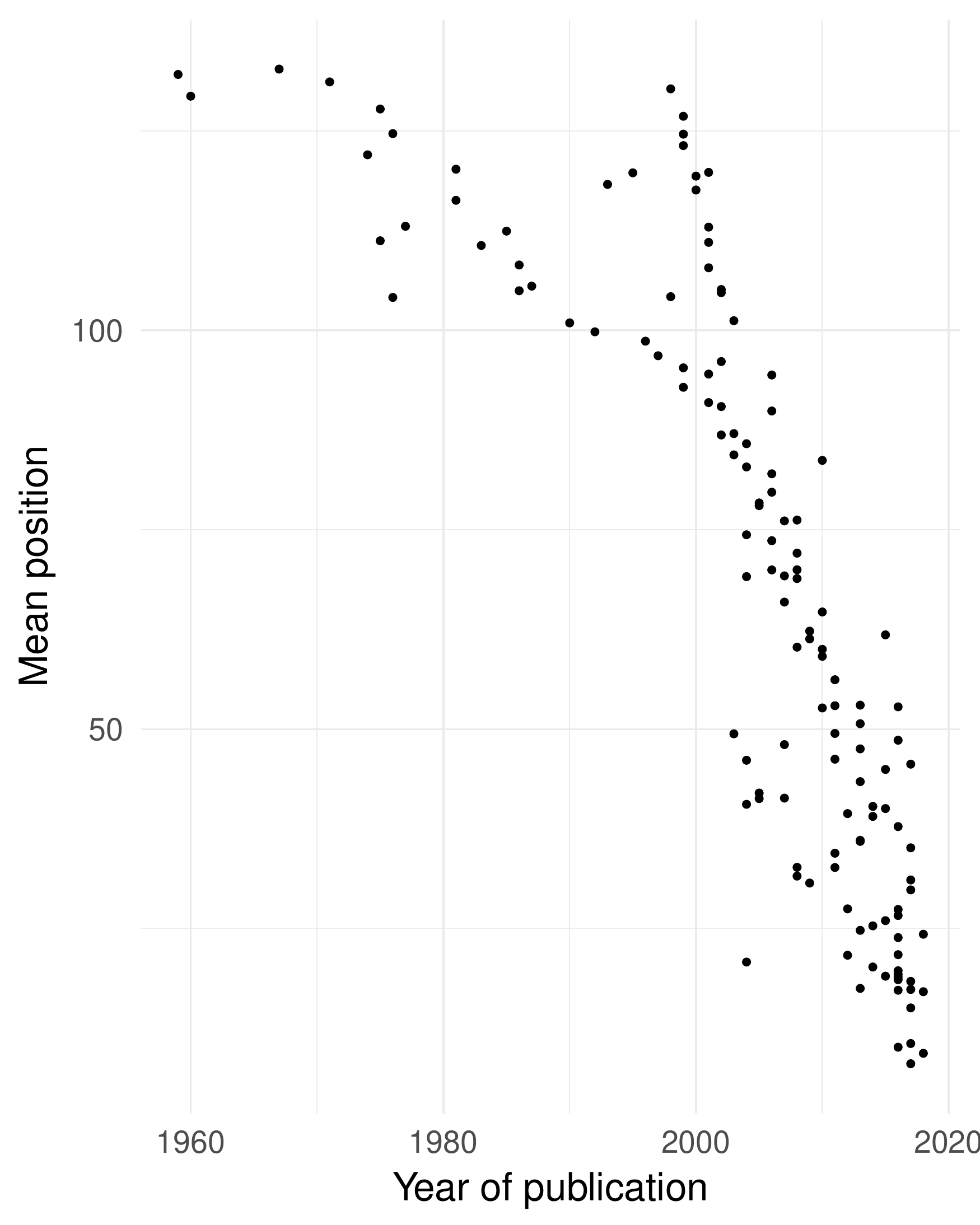} 

}

\caption[Posterior mean of position in $\o$ against year of publication, for $K$ = 3 (top left), 4 (top right), 5 (bottom left) and 6 (bottom right)]{Posterior mean of position in $\o$ against year of publication, for $K$ = 3 (top left), 4 (top right), 5 (bottom left) and 6 (bottom right).}\label{fig.plot_rgs_i_vs_year}
\end{figure}

\end{knitrout}

\section{Discussion} \label{sect.discuss}
In this article we presented a social network analysis on a citation network data set, in contrast to the collaboration networks that are more commonly investigated from the social network point of view. We argue that clustering analysis, in particular the application of one kind of the SBMs, is appropriate and useful for getting insights on the different groups of articles in the literature. Not only is our proposed model able to recover the three main groups, but it also infers how different the mixed memberships of various reviews are. While the results are potentially useful for seeing a larger picture of the literature in social network analysis, the model can be applied to any other and possibly much larger citation networks, or other kinds of DAGs such as software dependencies.

Connecting two articles using whether one cites the other comes with the issue that some relevant articles may be left out because they have not cited and been cited by others in the data. Not being connected to the giant component may come with huge uncertainty in the inference. One way of mitigating this issue is to consider the number of co-citations between two papers. Essentially the network will no longer be a DAG but an undirected graph with weighted edges. Constructing the network in this way is more likely to lead to a connected graph with potentially more information on the clustering of the articles.

The inference algorithm could be modified to improve efficiency. Instead of the usual way of integrating $\C$ and $\D$ out to obtain the collapsed Gibbs sampler, which might not be computationally more efficient the regular Gibbs sampler, one can consider integrating $\Z$ out in the likelihood, for a few reasons. Firstly, the sheer number of latent variables makes $\Z$ less interpretable than $\D$, $\C$, $\a$ and $\o$. Secondly, as $K$ is usually much smaller than $n$ if one really wants to discover communities, integrating $\Z$ out, which means summing over all $K^2$ possibilities each pair of latent variables $(\Zstar[pq],\Zstar[qp])$ can take, will not add much to the computational cost. Finally, the resulting likelihood will be continuous over $\D$ and $\C$, rendering Langevin and Hamiltonian Monte Carlo methods possible.

Regarding the topological order, the low dependence of $\o$ on the choice of $K$ may prompt the question that if $\o$ can be integrated out from the joint posterior too, or if the marginal posterior of $\o$ can be derived explicitly, although this does not look straightforward at all according to \eqref{eqn.lik_inf_joint}. However, if the model is also modified according to the co-citation idea aforementioned, this issue will not arise as the graph becomes undirected.

One more insight that could be drawn on the posterior of $\o$ is from its scatterplot against the chronological order. Departure of the mean position from the smoothed line may indicate higher or lower than average influence on the literature. This, of course, should be based on the assumption that the data is representative of the literature in the field of interest.

\section*{Abbreviations}
\textbf{MMSBM}: mixed membership stochastic block model; \textbf{DAG}: directed acyclic graph; \textbf{ERGM}: exponential random graph model; \textbf{SBM}: stochastic block model; \textbf{MCMC}: Markov chain Monte Carlo

\section*{Declarations}
\subsection*{Funding/Acknowledgements}
This research was funded by the Engineering and Physical Sciences Research Council (EPSRC) grant DERC: Digital Economy Research Centre (EP/M023001/1).

\subsection*{Availability of Data and Materials}
Data supporting this publication is openly available under an `Open Data Commons Open Database License'. Additional metadata are avaiable at:\\http://dx.doi.org/10.17634/141304-14. Please contact Newcastle Research Data Service at rdm@ncl.ac.uk for access instructions.

\subsection*{Competing Interests}
The authors declare that they have no competing interests.

\subsection*{Authors' Contributions}
CL and DJW developed the model. CL collected the data, performed the analyses, and wrote the paper. Both authors reviewed and approved the manuscript.

\bibliographystyle{agsm}
\bibliography{ref_nr_no_url}

\renewcommand{\listfigurename}{Figure Legend}
\listoffigures

\begin{center}
{\large\bf APPENDICES}
\end{center}
\appendix

\section{Regular Gibbs sampler} \label{sect.appendix_rgs}
\subsection*{Updating $\C$}
According to \eqref{eqn.lik_inf_joint}, the conditional posterior of $\C$ given $\Y$, $\Z$, $\o$ (and $\D$ and $\a$) can be rearranged to become
\begin{align}
  &\pi(\C|\Y,\Z,\D,\a,\o)\nonumber\\
  \propto~&\Prod[n-1]{p=1}\Prod[n]{q=p+1}\C[{\Zstar[pq]\Zstar[qp]}]^{\Ystar[pq]}\left(1-\C[{\Zstar[pq]\Zstar[qp]}]\right)^{\left(1-\Ystar[pq]\right)}\times\Prod[K]{i=1}\Prod[K]{j=1}\C[ij]^{\A[ij]-1}\left(1-\C[ij]\right)^{\B[ij]-1}\nonumber\\
  =~&\Prod[K]{i=1}\Prod[K]{j=1}\C[ij]^{\mathbf{E}_{ij}}(1-\C[ij])^{\mathbf{F}_{ij}}\times\Prod[K]{i=1}\Prod[K]{j=1}\C[ij]^{\A[ij]-1}\left(1-\C[ij]\right)^{\B[ij]-1},\nonumber\\ \intertext{where}
  &\mathbf{E}_{ij}=\sum^{n-1}_{p=1}\sum^{n}_{q=p+1}\left[\mathbf{1}\{\Zstar[pq]=i,\Zstar[qp]=j\}\Ystar[pq]\right],\nonumber\\
  &\mathbf{F}_{ij}=\sum^{n-1}_{p=1}\sum^{n}_{q=p+1}\left[\mathbf{1}\{\Zstar[pq]=i,\Zstar[qp]=j\}\left(1-\Ystar[pq]\right)\right].\nonumber
\end{align}
This leads to the Gibbs step for $\C[ij],1\leq i,j\leq K$:
\begin{align}
  \C[ij]|\cdots\sim\text{Beta}\left(\mathbf{E}_{ij}+\A[ij],\mathbf{F}_{ij}+\B[ij]\right).\nonumber
\end{align}

\subsection*{Updating $\D$}
For $p,q=1,2,\ldots,n~(p\neq q)$ and $i=1,2,\ldots,K$, $\Dstar[{p\Zstar[pq]}]$ can be written as $\Prod[K]{i=1}{\Dstar[pi]}^{\one{\Zstar[pq]=i}}$. Therefore, according to \eqref{eqn.lik_inf_joint}, the conditional posterior of $\D$ given $\Y$, $\Z$, $\o$, $\a$ (and $\C$) can be rearranged to become
\begin{align}
  &\pi(\D|\Y,\Z,\C,\a,\o)\nonumber\\
  \propto~&\Prod{p<q}\Dstar[{p\Zstar[pq]}]\Dstar[{q\Zstar[qp]}]\times\Prod[n]{p=1}\left[\one{\sum_{i=1}^{K}\Dstar[pi]=1}\Prod[K]{i=1}~\Dstar[pi]{}^{(\a-1)}\right]\nonumber\\
  =~&\Prod{p<q}\Dstar[{p\Zstar[pq]}]\times\Prod{p>q}\Dstar[{p\Zstar[pq]}]\times\Prod[n]{p=1}\left[\one{\sum_{i=1}^{K}\Dstar[pi]=1}\Prod[K]{i=1}~\Dstar[pi]{}^{(\a-1)}\right]\nonumber\\
  =~&\Prod{p\neq q}\Prod[K]{i=1}\Dstar[pi]{}^{\one{\Zstar[pq]=i}}\times\Prod[n]{p=1}\left[\one{\sum_{i=1}^{K}\Dstar[pi]=1}\Prod[K]{i=1}~\Dstar[pi]{}^{(\a-1)}\right]\nonumber\\
  =~&\Prod[n]{p=1}\Prod[K]{i=1}\Dstar[pi]{}^{\sum_{q\neq p}\one{\Zstar[pq]=i}}\times\Prod[n]{p=1}\left[\one{\sum_{i=1}^{K}\Dstar[pi]=1}\Prod[K]{i=1}~\Dstar[pi]{}^{(\a-1)}\right]\nonumber\\
  =~&\Prod[n]{p=1}\left[\one{\sum_{i=1}^{K}\Dstar[pi]=1}\Prod[K]{i=1}\Dstar[pi]{}^{\left(\a_{pi}-1\right)}\right]\nonumber\\ \intertext{where}
  &\a_{pi}=~\a+\sum_{q\neq p}\one{\Zstar[pq]=i}=\a+\sum_{q=1}^{n}\one{\Zstar[pq]=i}.\nonumber
\end{align}
The last line is due to that $\Zstar[pp]$ can never be equal to $i$ as it is fixed to $-1$. Now, we have the Gibbs step for $\dstar[p]=\left(\Dstar[p1]~\Dstar[p2]~\cdots~\Dstar[pK]\right)^{T}$, $p=1,2,\ldots,n$:
\begin{align}
  \dstar[p]|\cdots\sim\text{Dirichlet}\left(\left(\a_{p1}~\a_{p2}~\cdots~\a_{pK}\right)^T\right).\nonumber
\end{align}

\subsection*{Updating $\a$}
The conditional posterior density of $\a$ given $\D$ (and $\Y$, $\Z$, $\C$ and $\o$) is
\begin{align}
  \pi(\a|\Y,\Z,\D,\C,\o)\propto~&\frac{\Gamma(K\a)^{n}}{\Gamma(\a)^{nK}}\times\left[\Prod[n]{p=1}\Prod[K]{i=1}{\Dstar[pi]}\right]^{\a}\times\a^{a-1}e^{-b\a}\one{\a>0}.\nonumber
\end{align}
To update $\a$ via a Metropolis step, we propose $\a^{*}$ from $\text{N}\left(\a,s_{\a}^2\right)$, where $s_{\a}>0$ is the proposal standard deviation, and accept $\a^{*}$ with probability 
\begin{align}
  \mathlarger1\wedge\frac
    {\displaystyle\frac{\Gamma(K\a^{*})^{n}}{\Gamma(\a^{*})^{nK}}\times\left[\Prod[n]{p=1}\Prod[K]{i=1}{\Dstar[pi]}\right]^{\a^{*}}\times\left(\a^{*}\right)^{a-1}e^{-b\a^{*}}\one{\a^{*}>0}}
    {\displaystyle\frac{\Gamma(K\a)^{n}}{\Gamma(\a)^{nK}}\times\left[\Prod[n]{p=1}\Prod[K]{i=1}{\Dstar[pi]}\right]^{\a}\times\a^{a-1}e^{-b\a}\one{\a>0}}
    .\nonumber
\end{align}

\subsection*{Updating $\Zstar[pq]~(p<q)$}
The conditional posterior of $\Zstar[pq]$ conditional on everything else is
\begin{align}
  \pi\left(\Zstar[pq]|\Y,\Zstar[{-pq}],\C,\D,\a,\o\right)
  \propto~&\C[{\Zstar[pq]\Zstar[qp]}]^{\Ystar[pq]}\left(1-\C[{\Zstar[pq]\Zstar[qp]}]\right)^{1-\Ystar[pq]}\Dstar[{p\Zstar[pq]}],\nonumber
\end{align}
where $\Zstar[{-pq}]$ means all of $\Zstar$ except $\Zstar[pq]$. Therefore, we update $\Zstar[pq]$ with a Gibbs step where
\begin{align}
  \Pr\left(\Zstar[pq]=i|\cdots\right)=\frac
          {\C[{i\Zstar[pq]}]^{\Ystar[pq]}\left(1-\C[{i\Zstar[pq]}]\right)^{1-\Ystar[pq]}\Dstar[{pi}]}
          {\displaystyle\sum_{j=1}^{K}\C[{j\Zstar[pq]}]^{\Ystar[pq]}\left(1-\C[{j\Zstar[pq]}]\right)^{1-\Ystar[pq]}\Dstar[{pj}]},\nonumber
\end{align}
for $i=1,2,\ldots,K$. This update is essentially a reweighting according to the $\Zstar[qp]$-\textit{th} column of $\C$ and the $p$-\textit{th} row of $\Dstar$.

\subsection*{Updating $\Zstar[qp]~(p<q)$}
The conditional posterior of $\Zstar[qp]$ conditional on everthing else is
\begin{align}
  \pi\left(\Zstar[qp]|\Y,\Zstar[{-qp}],\C,\D,\a,\o\right)
  \propto~&\C[{\Zstar[pq]\Zstar[qp]}]^{\Ystar[pq]}\left(1-\C[{\Zstar[pq]\Zstar[qp]}]\right)^{1-\Ystar[pq]}\Dstar[{q\Zstar[qp]}].\nonumber
\end{align}
Therefore, we update $\Zstar[qp]$ with a Gibbs step where
\begin{align}
  \Pr\left(\Zstar[qp]=j|\cdots\right)=\frac
          {\C[{\Zstar[pq]j}]^{\Ystar[pq]}\left(1-\C[{\Zstar[pq]j}]\right)^{1-\Ystar[pq]}\Dstar[{qj}]}
          {\displaystyle\sum_{i=1}^{K}\C[{\Zstar[pq]i}]^{\Ystar[pq]}\left(1-\C[{\Zstar[pq]i}]\right)^{1-\Ystar[pq]}\Dstar[{qi}]},\nonumber
\end{align}
for $j=1,2,\ldots,K$. This update is essentially a reweighting according to the $\Zstar[pq]$-\textit{th} row of $\C$ and the $q$-\textit{th} row of $\Dstar$.

\subsection*{Updating $\o$ (and the star matrices)}
To update $\o$, we will carry out a number of proposals to swap pairs of adjacent elements in $\o$. We will not adopt the random insertion method by \cite{bks06} because $\Ystar$ is more likely to be no longer upper triangular if the elements in $\o$ move ``radically''. The updating scheme is as follows:
\begin{enumerate}
  \item Select an index, denoted by $p$, from $\{1,2,\ldots,n\}$ at random uniformly. If $p=1$, set another index, denoted by $q$, to 2. If $p=n$, set $q=n-1$. Otherwise set $q=p-1$ and $q=p+1$ with probability 0.5 each.
  \item If $\Ystar[\min(p,q),\max(p,q)]=1$, keep $\o$ unchanged, as swapping $\o[p]$ and $\o[q]$ will make $\Ystar$ no longer upper triangular and yield a zero likelihood. Otherwise propose to swap $\o[p]$ and $\o[q]$ according to the following five scenarios.
  \item \label{list.p1} If $p=1$, swap $\o[1]$ and $\o[2]$ with probability $1\wedge\displaystyle\frac{\left(1-\C[{\Zstar[21]\Zstar[12]}]\right)}{2\left(1-\C[{\Zstar[12]\Zstar[21]}]\right)}$.
  \item \label{list.pn} If $p=n$, swap $\o[n]$ and $\o[n-1]$ with probability $1\wedge\displaystyle\frac{\left(1-\C[{\Zstar[{(n-1)n}]\Zstar[{n(n-1)}]}]\right)}{2\left(1-\C[{\Zstar[{n(n-1)}]\Zstar[{(n-1)n}]}]\right)}$.
  \item \label{list.q1} If $q=1$, swap $\o[1]$ and $\o[2]$ with probability $1\wedge\displaystyle\frac{2\left(1-\C[{\Zstar[{12}]\Zstar[{21}]}]\right)}{\left(1-\C[{\Zstar[{21}]\Zstar[{12}]}]\right)}$.
  \item \label{list.qn} If $q=n$, swap $\o[n]$ and $\o[n-1]$ with probability $1\wedge\displaystyle\frac{2\left(1-\C[{\Zstar[{n(n-1)}]\Zstar[{(n-1)n}]}]\right)}{\left(1-\C[{\Zstar[{(n-1)n}]\Zstar[{n(n-1)}]}]\right)}$.
  \item \label{list.between} If $p$ and $q$ together do not belong to any of the four scenarios above, swap $\o[p]$ and $\o[q]$ with probability $1\wedge\displaystyle\frac{1-\C[{\Zstar[{qp}]\Zstar[{pq}]}]}{1-\C[{\Zstar[{pq}]\Zstar[{qp}]}]}$.
  \item Update $\Ystar$, $\Zstar$ and $\Dstar$ according to current $\Y$, $\Z$, $\D$ and $\o$.
\end{enumerate}

\section{Traceplots and posterior densities} \label{sect.appendix_plots}
The traceplots and posterior densities of $\a$ for different $K$ are plotted in Figure \ref{fig.plot_rgs_alpha_trace}. The traceplots and posterior densities of the first row of $\C$ for different $K$ are plotted in Figures \ref{fig.plot_rgs_C_trace1} to \ref{fig.plot_rgs_C_trace4}. The traceplots and posterior densities of $\D$ for a selected article \citep{abfx08} for different $K$ are plotted in Figures \ref{fig.plot_rgs_D_trace1} to \ref{fig.plot_rgs_D_trace4}. The traceplots and posterior histograms of the position of \cite{abfx08} in $\o$ for different $K$ are plotted in Figure \ref{fig.plot_rgs_o_trace}
\begin{knitrout}
\definecolor{shadecolor}{rgb}{0.969, 0.969, 0.969}\color{fgcolor}\begin{figure}[htbp!]

{\centering \includegraphics[width=0.98\linewidth]{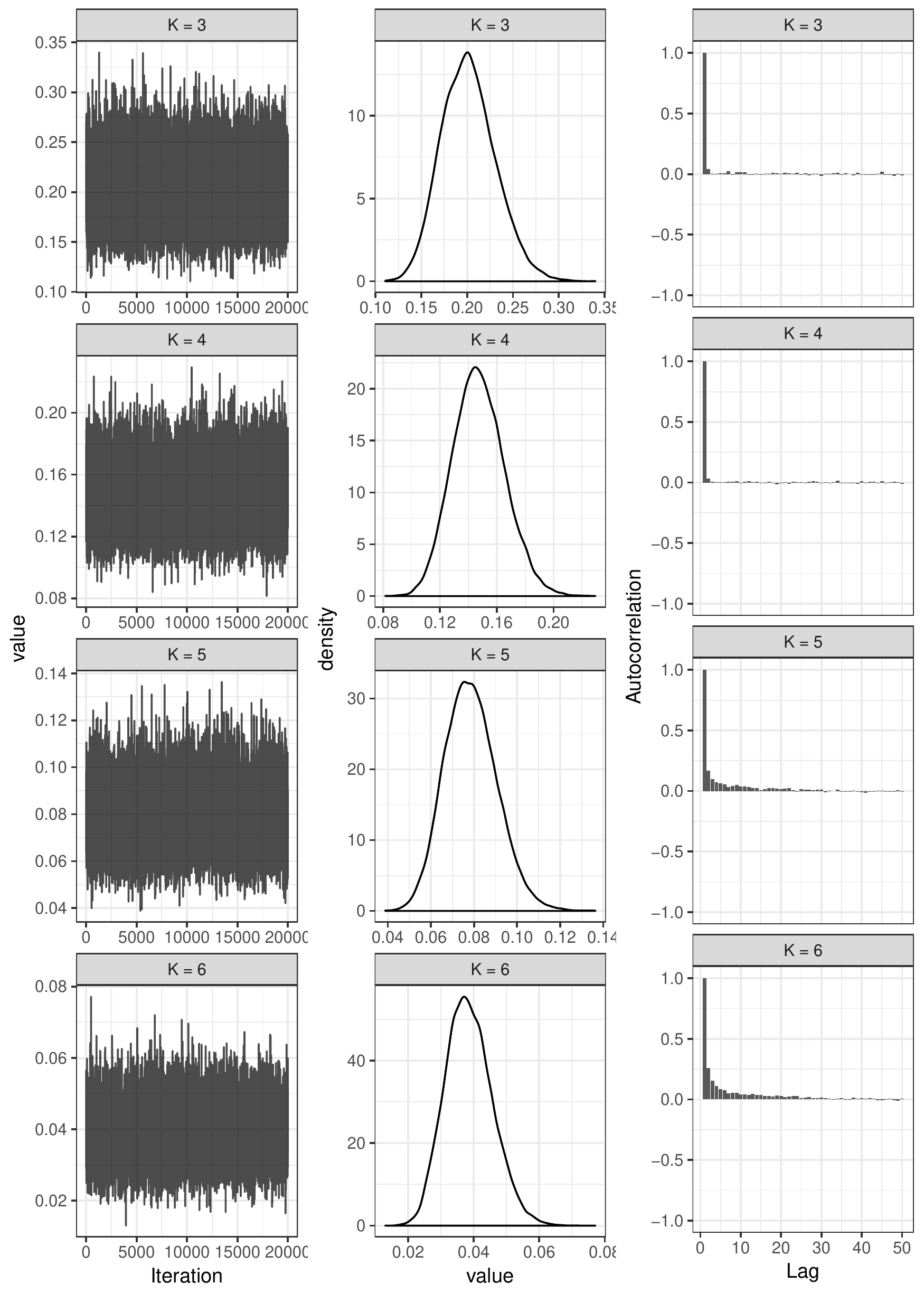} 

}

\caption[Traceplots (left), posterior densities (middle) and autocorrelation functions (ACFs, right) of $\a$ for $K$ = 3, 4, 5 and 6]{Traceplots (left), posterior densities (middle) and autocorrelation functions (ACFs, right) of $\a$ for $K$ = 3, 4, 5 and 6.}\label{fig.plot_rgs_alpha_trace}
\end{figure}

\end{knitrout}

\begin{knitrout}
\definecolor{shadecolor}{rgb}{0.969, 0.969, 0.969}\color{fgcolor}\begin{figure}[htbp!]

{\centering \includegraphics[width=0.98\linewidth]{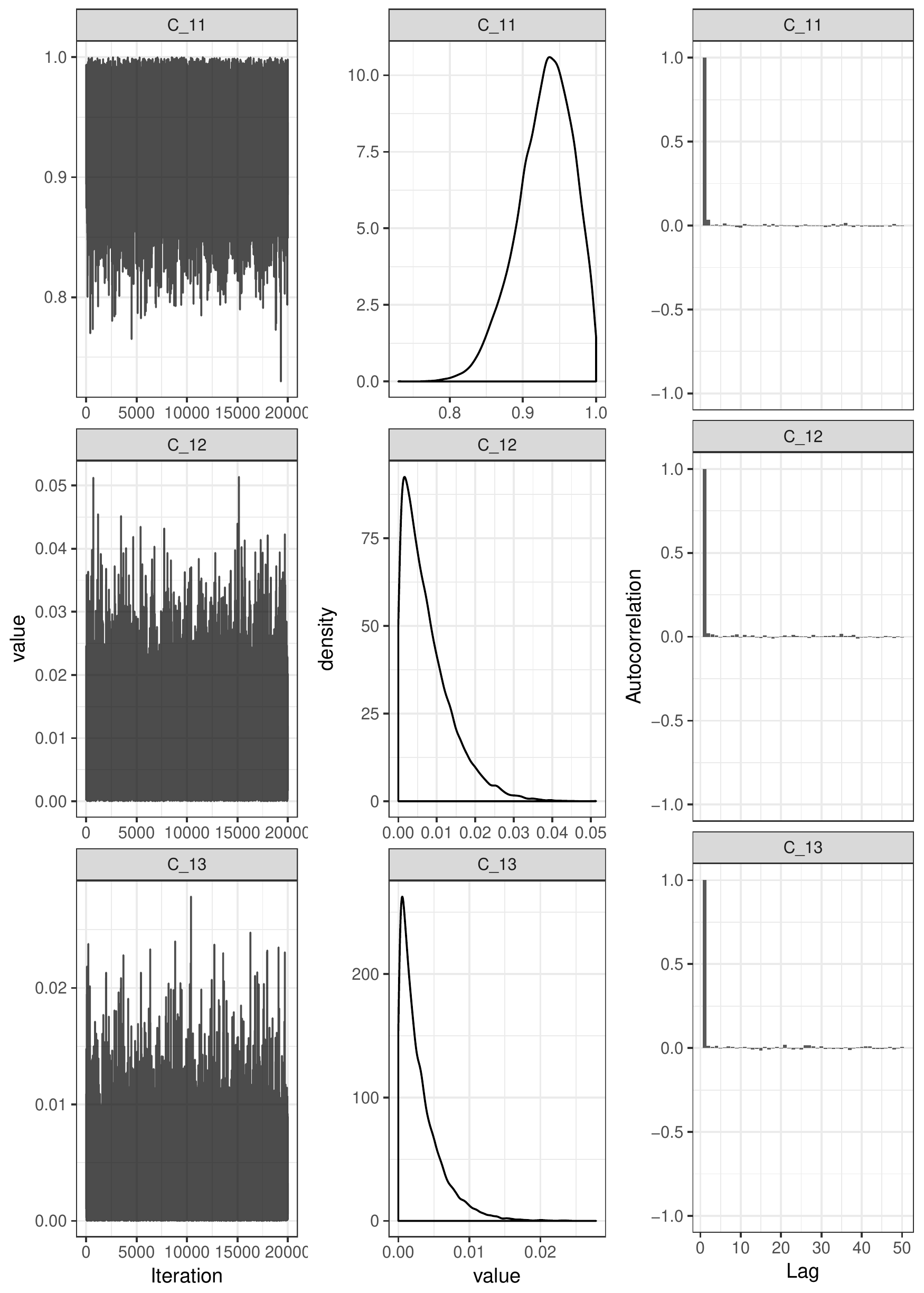} 

}

\caption[Traceplots (left), posterior densities (middle) and ACFs (right) of the first row of $\C$ for $K$ = 3]{Traceplots (left), posterior densities (middle) and ACFs (right) of the first row of $\C$ for $K$ = 3.}\label{fig.plot_rgs_C_trace1}
\end{figure}

\begin{figure}[htbp!]

{\centering \includegraphics[width=0.98\linewidth]{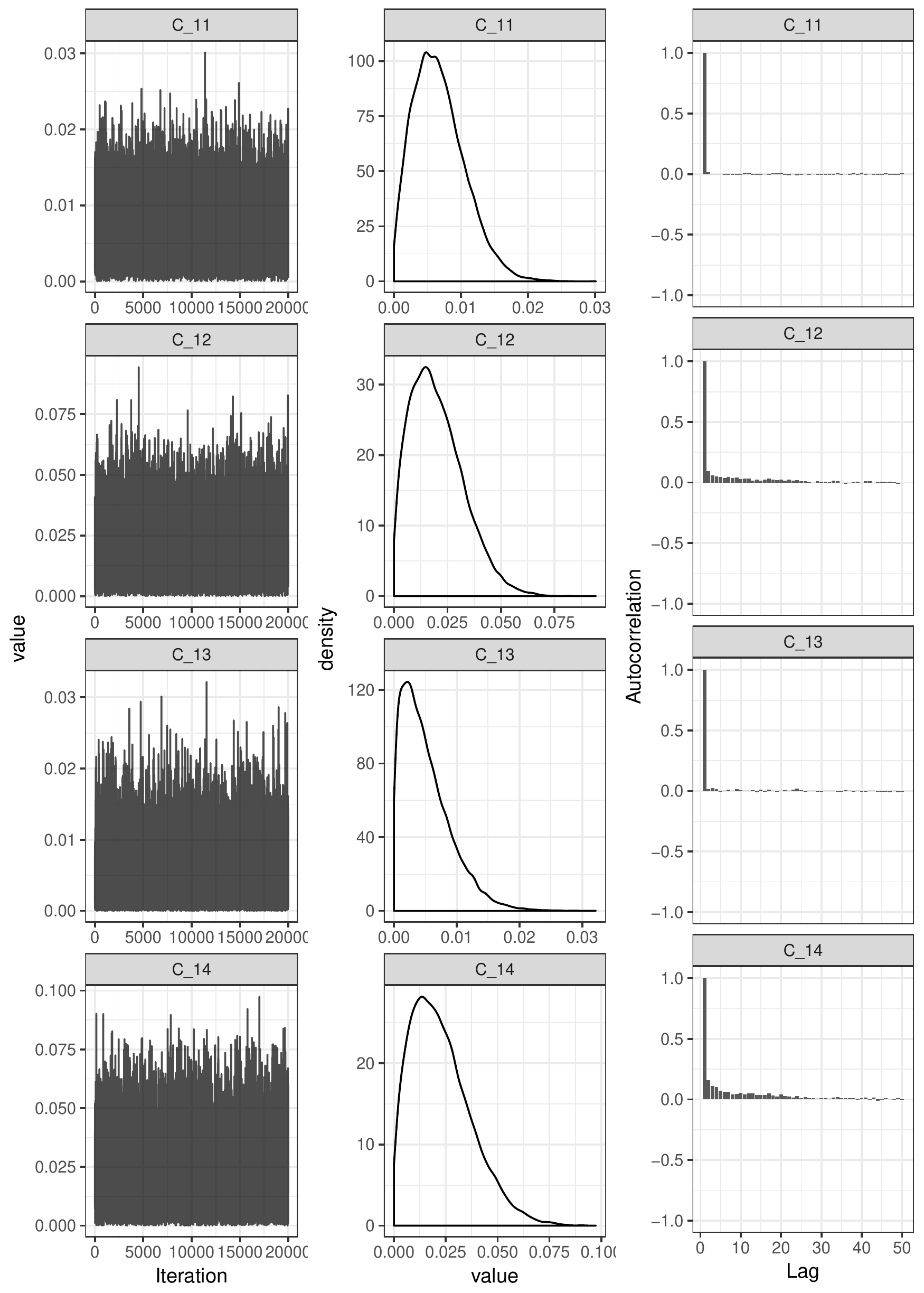} 

}

\caption[Traceplots (left), posterior densities (middle) and ACFs (right) of the first row of $\C$ for $K$ = 4]{Traceplots (left), posterior densities (middle) and ACFs (right) of the first row of $\C$ for $K$ = 4.}\label{fig.plot_rgs_C_trace2}
\end{figure}

\begin{figure}[htbp!]

{\centering \includegraphics[width=0.98\linewidth]{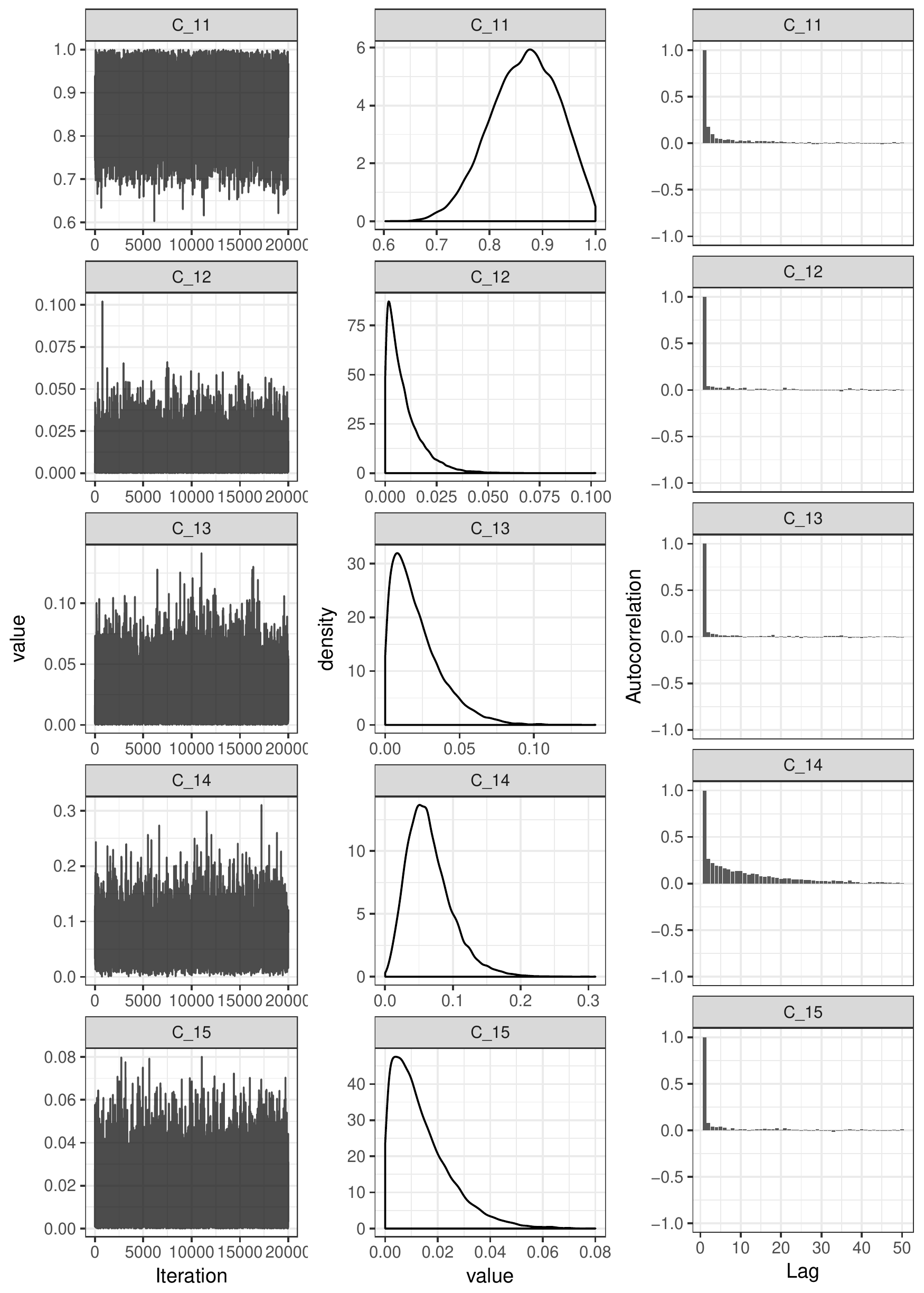} 

}

\caption[Traceplots (left), posterior densities (middle) and ACFs (right) of the first row of $\C$ for $K$ = 5]{Traceplots (left), posterior densities (middle) and ACFs (right) of the first row of $\C$ for $K$ = 5.}\label{fig.plot_rgs_C_trace3}
\end{figure}

\begin{figure}[htbp!]

{\centering \includegraphics[width=0.98\linewidth]{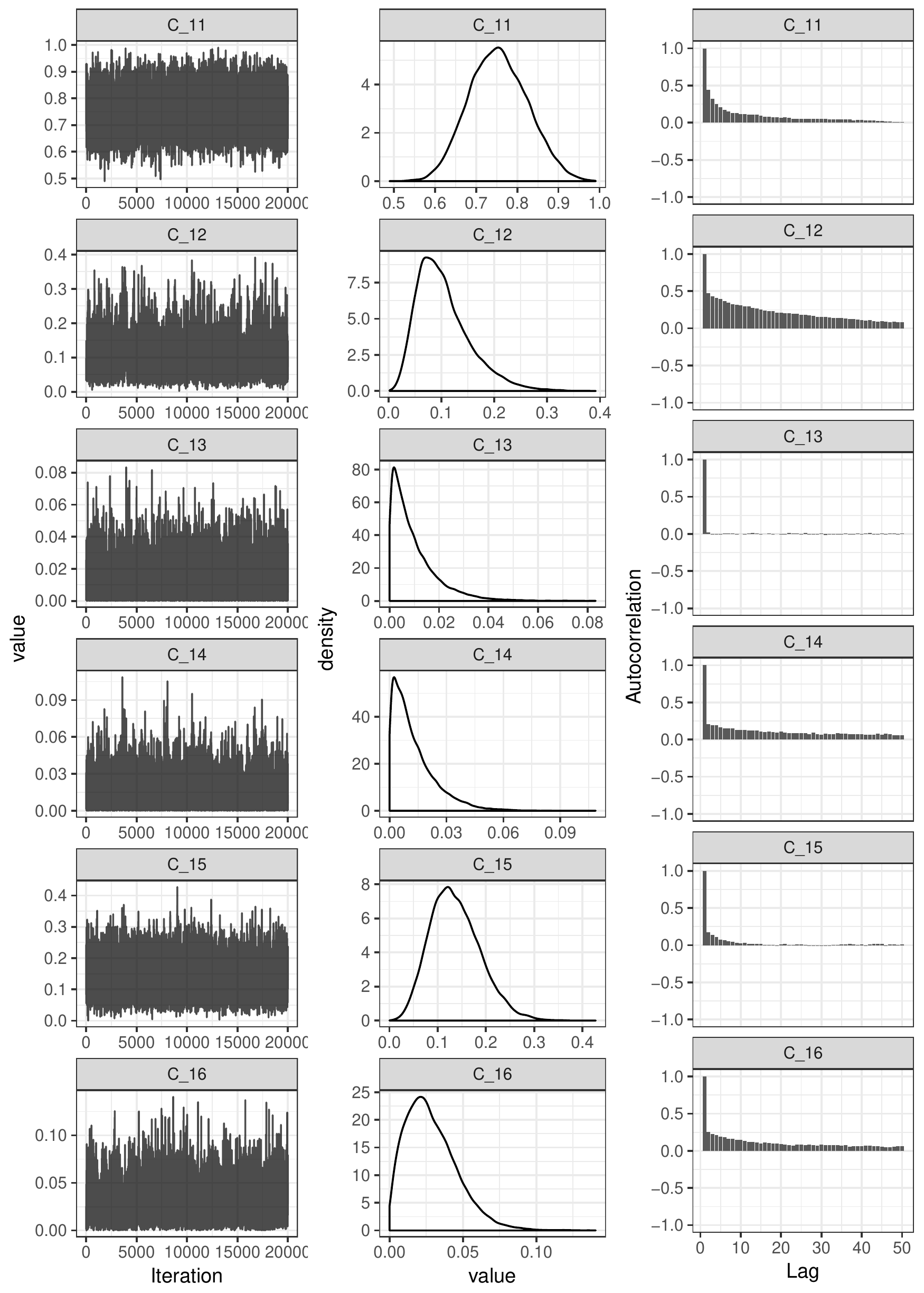} 

}

\caption[Traceplots (left), posterior densities (middle) and ACFs (right) of the first row of $\C$ for $K$ = 6]{Traceplots (left), posterior densities (middle) and ACFs (right) of the first row of $\C$ for $K$ = 6.}\label{fig.plot_rgs_C_trace4}
\end{figure}

\end{knitrout}

\begin{knitrout}
\definecolor{shadecolor}{rgb}{0.969, 0.969, 0.969}\color{fgcolor}\begin{figure}[htbp!]

{\centering \includegraphics[width=0.95\linewidth]{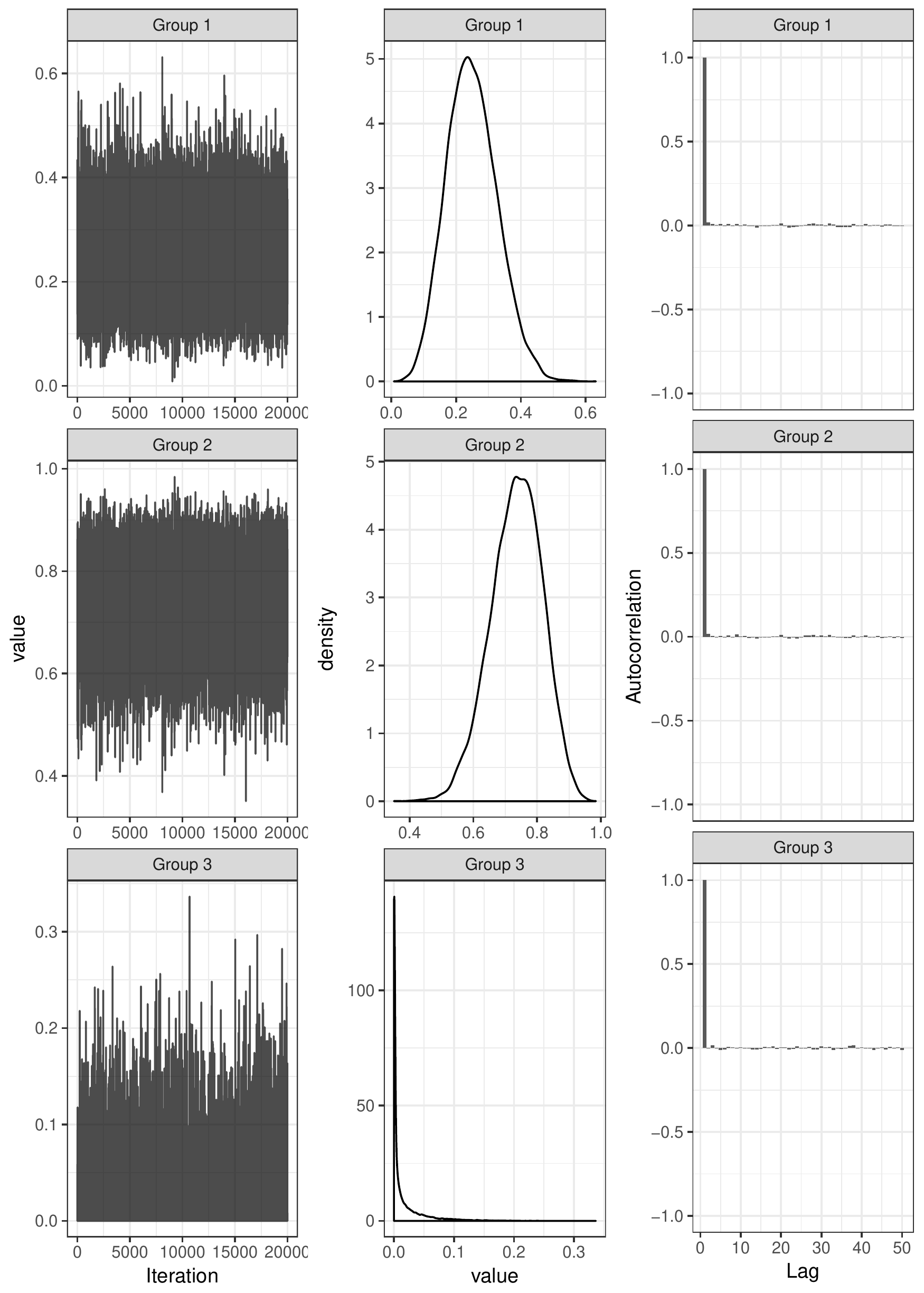} 

}

\caption[Traceplots (left), posterior densities (middle) and ACFs (right) of the group memberships of \cite{abfx08} for $K$ = 3]{Traceplots (left), posterior densities (middle) and ACFs (right) of the group memberships of \cite{abfx08} for $K$ = 3.}\label{fig.plot_rgs_D_trace1}
\end{figure}

\begin{figure}[htbp!]

{\centering \includegraphics[width=0.95\linewidth]{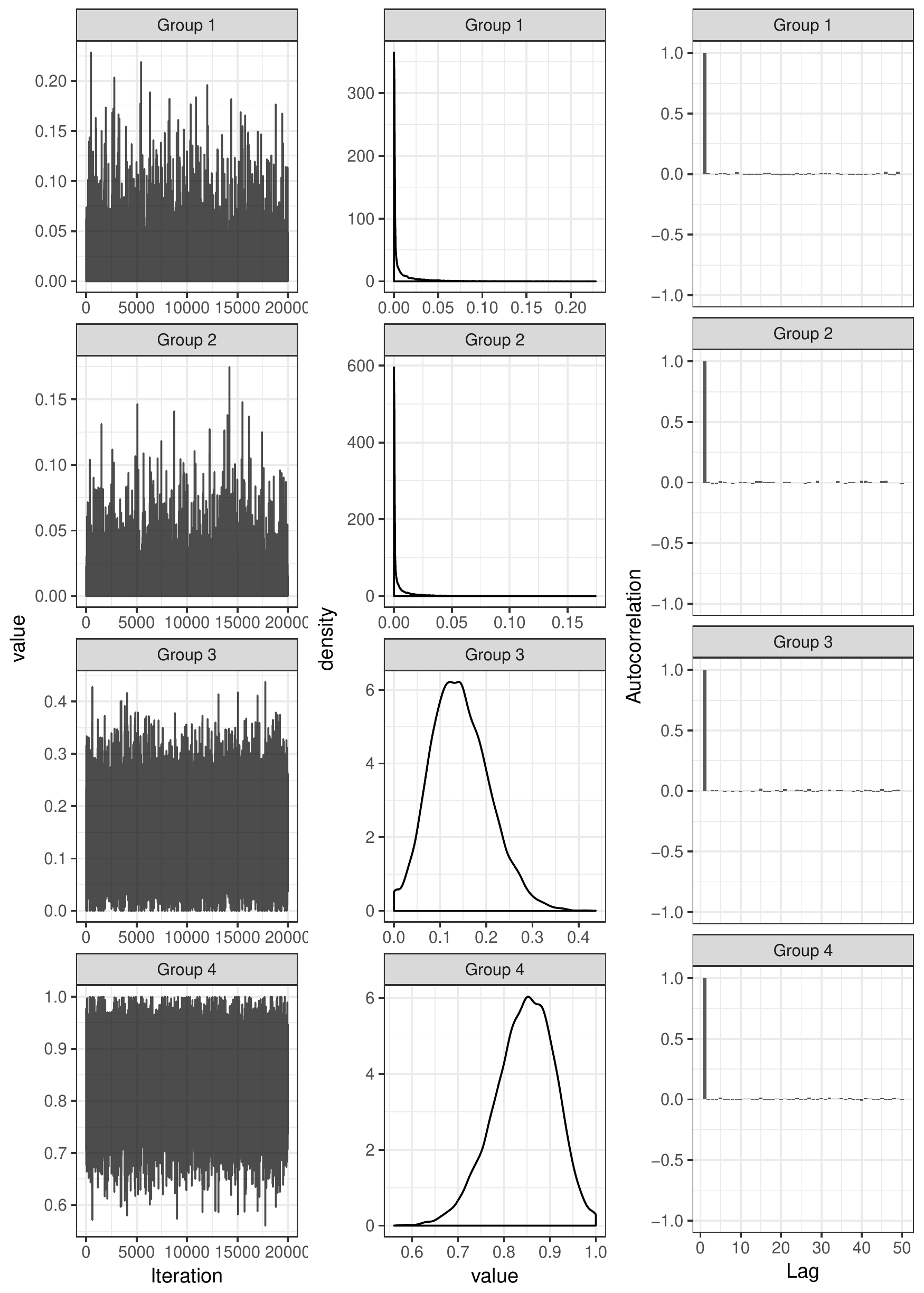} 

}

\caption[Traceplots (left), posterior densities (middle) and ACFs (right) of the group memberships of \cite{abfx08} for $K$ = 4]{Traceplots (left), posterior densities (middle) and ACFs (right) of the group memberships of \cite{abfx08} for $K$ = 4.}\label{fig.plot_rgs_D_trace2}
\end{figure}

\begin{figure}[htbp!]

{\centering \includegraphics[width=0.95\linewidth]{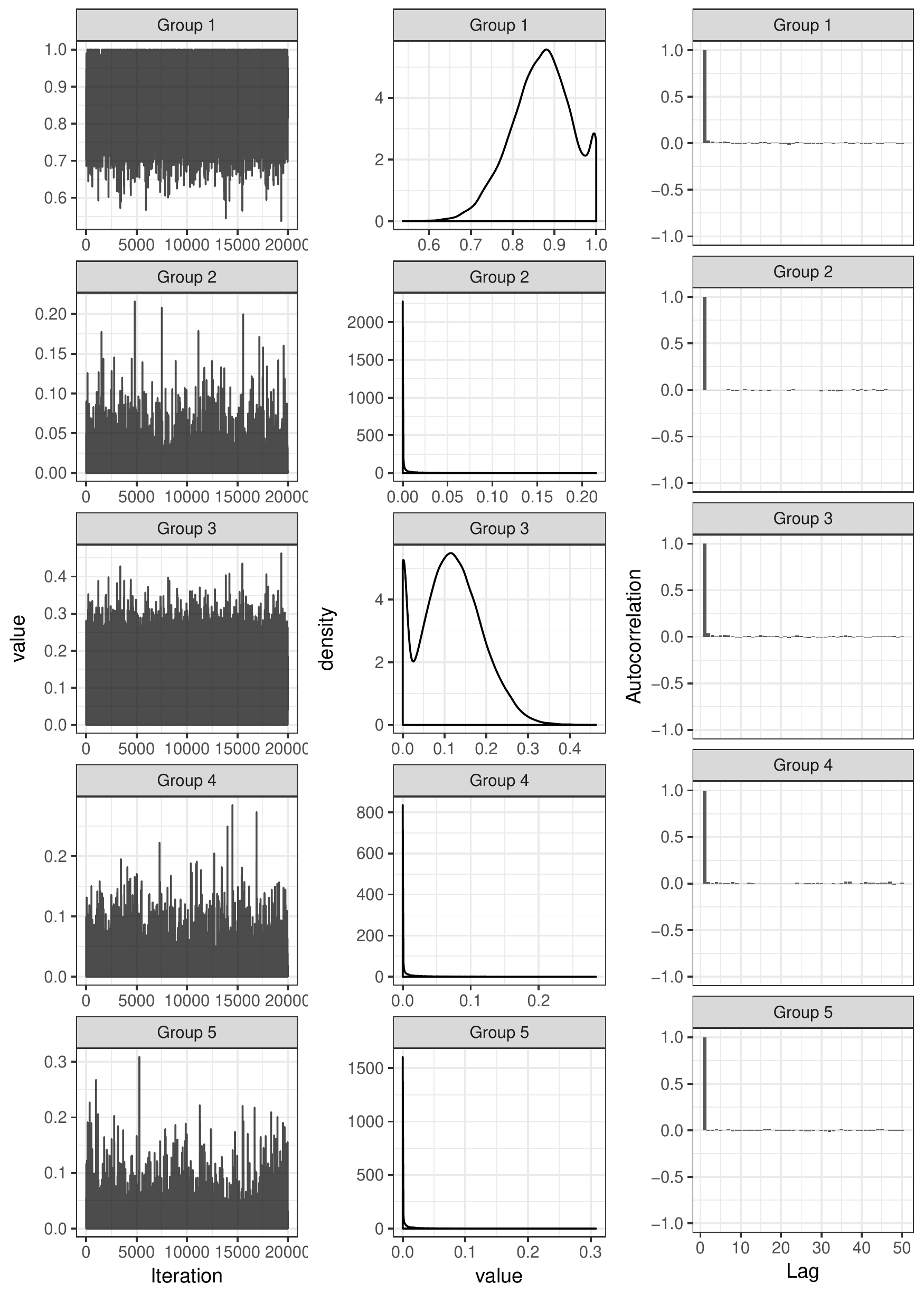} 

}

\caption[Traceplots (left), posterior densities (middle) and ACFs (right) of the group memberships of \cite{abfx08} for $K$ = 5]{Traceplots (left), posterior densities (middle) and ACFs (right) of the group memberships of \cite{abfx08} for $K$ = 5.}\label{fig.plot_rgs_D_trace3}
\end{figure}

\begin{figure}[htbp!]

{\centering \includegraphics[width=0.95\linewidth]{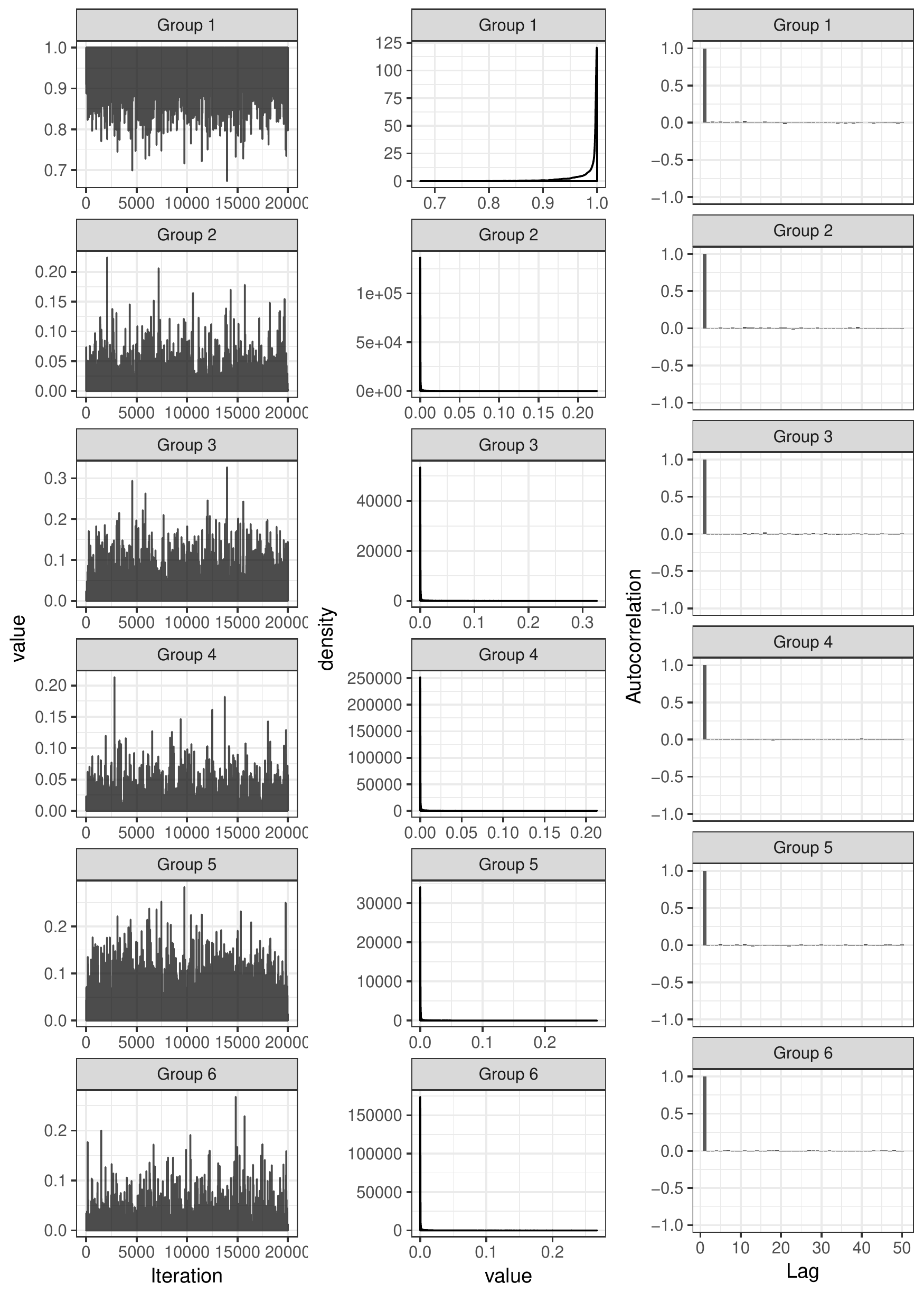} 

}

\caption[Traceplots (left), posterior densities (middle) and ACFs (right) of the group memberships of \cite{abfx08} for $K$ = 6]{Traceplots (left), posterior densities (middle) and ACFs (right) of the group memberships of \cite{abfx08} for $K$ = 6.}\label{fig.plot_rgs_D_trace4}
\end{figure}

\end{knitrout}

\begin{knitrout}
\definecolor{shadecolor}{rgb}{0.969, 0.969, 0.969}\color{fgcolor}\begin{figure}[htbp!]

{\centering \includegraphics[width=0.95\linewidth]{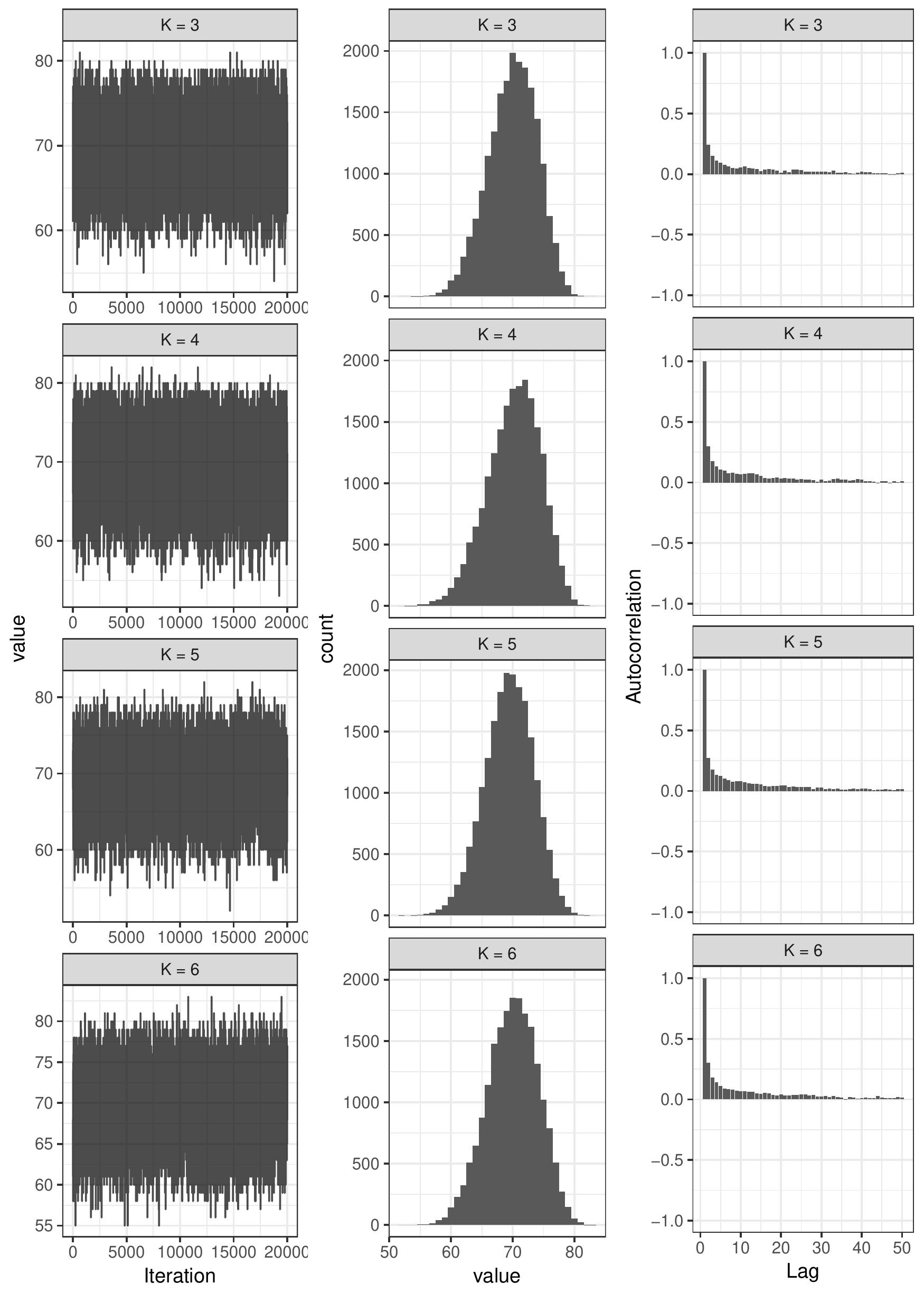} 

}

\caption[Traceplots (left), posterior histograms (middle) and ACFs (right) of the positions of \cite{abfx08} in $\o$ for $K$ = 3, 4, 5 and 6]{Traceplots (left), posterior histograms (middle) and ACFs (right) of the positions of \cite{abfx08} in $\o$ for $K$ = 3, 4, 5 and 6.}\label{fig.plot_rgs_o_trace}
\end{figure}

\end{knitrout}

\end{document}